\documentclass[conference]{IEEEtran}
\IEEEoverridecommandlockouts

\usepackage{amsmath, mathtools, amsthm, amssymb}
\usepackage{proof}
\usepackage{stmaryrd}

\usepackage[table]{xcolor}
\usepackage{colortbl}
\usepackage{hyperref}
\hypersetup{hidelinks}
\usepackage{tcolorbox}
\usepackage{tabularx}
\usepackage{booktabs}
\usepackage{pifont}
\usepackage{enumitem}
\usepackage{threeparttable}
\usepackage{multicol}
\usepackage{graphicx}
\usepackage{rotating}
\usepackage{makecell}
\usepackage{multirow}
\usepackage{cite}

\tcbuselibrary{skins}

\definecolor{xhs_blue}{HTML}{364FC7}
\definecolor{xhs_blue_dark}{HTML}{081C59}
\definecolor{xhs_blue_light}{HTML}{EDF2FF}

\definecolor{xhs_red}{HTML}{c13c33}
\definecolor{xhs_green}{HTML}{3c5840}

\definecolor{xhs_all}{HTML}{9f1f31}
\definecolor{xhs_ex}{HTML}{2885a9}

\NewTotalTColorBox{\BlueQuote}{ +m +m }{
    notitle,
    colback=xhs_blue_light,
    colbacklower=white,
    frame hidden,
    boxrule=0pt,
    bicolor,
    sharp corners,
    borderline west={2.5pt}{0pt}{xhs_blue}
}{
    \textcolor{xhs_blue}{ \textbf{#1} }
    #2
}

\newcommand{\funcText}[1]{{\ttfamily #1}}

\newcommand{\ruleNameText}[1]{{\ttfamily #1}}

\newcommand{\TFact}[1]{{\ttfamily #1}}
\newcommand{\TLemma}[1]{{\ttfamily #1}}

\newcommand{\TFacts}[1]{\lbrack\ #1 \ \rbrack}
\newcommand{\TEmpActFact}[0]{-\hspace{-0.45em}\rightarrow}

\newcommand{\TPairN}[1]{$\langle$\text{\ttfamily #1}$\rangle$}

\newcommand{\TLetS}[0]{\text{\Tterm{let}} \hspace{0.5em}}
\newcommand{\TInS}[0]{\text{\Tterm{in}} \hspace{0.5em}}

\newcommand{\TAll}[0]{\text{\textcolor{xhs_all}{\TLemma{All}}}}
\newcommand{\TEx}[0]{\text{\textcolor{xhs_ex}{\TLemma{Ex}}}}
\newcommand{\TNot}[0]{\text{\TLemma{not}}}

\newcommand{\Tterm}[1]{{\ttfamily #1}}

\newcommand{\Tfunction}[1]{{\ttfamily #1}}
\newcommand{\Tamarin}[0]{\textsc{Tamarin}}

\newcommand{\TFactTerm}[2]{{\ttfamily #1}$(\,$\Tterm{#2}$\,)$}

\newcolumntype{S}{>{\centering\arraybackslash}m{3cm}@{*}}

\begin{document}

\title{Small Leaks Sink Great Ships:\\ Automated Analysis of Protocols That Tolerate Low-Entropy Keys}

\author{
\IEEEauthorblockN{
Yongkang Xiao\IEEEauthorrefmark{1},
Jing Chen\IEEEauthorrefmark{1}\IEEEauthorrefmark{3}\thanks{\IEEEauthorrefmark{3}Corresponding authors: Jing Chen and Min Shi.},
Min Shi\IEEEauthorrefmark{1}\IEEEauthorrefmark{3},
Kun He\IEEEauthorrefmark{1},
Mei Wang\IEEEauthorrefmark{2},
Ruiying Du\IEEEauthorrefmark{1}}
\IEEEauthorblockA{
\IEEEauthorrefmark{1}School of Cyber Science and Engineering, Wuhan University\\
\IEEEauthorrefmark{2}School of Cyber Science and Technology, and the State Key Laboratory of\\
Cryptography and Digital Economy Security, Shandong University\\
\{xiaoyongkang, chenjing, itachi, hekun, duraying\}@whu.edu.cn\\
wangmeiz@sdu.edu.cn}
}

\makeatletter\def\@IEEEpubidpullup{6.5\baselineskip}\makeatother
\IEEEpubid{\parbox{\columnwidth}{
Network and Distributed System Security (NDSS) Symposium 2027\\
22--26 March 2027, Seoul, Republic of Korea\\
ISBN 978-1-970672-09-1\\
https://dx.doi.org/10.14722/ndss.2027.230474\\
www.ndss-symposium.org
}
\hspace{\columnsep}\makebox[\columnwidth]{}}

\maketitle

\begin{abstract}
Modern cryptographic protocols such as WPA2-PSK, Bluetooth Low Energy (BLE), and Bluetooth Classic (BC) tolerate low-entropy keys for user-friendliness, backward compatibility, or resource-constrained deployment, yet existing symbolic protocol verifiers rely on the perfect-encryption assumption and therefore cannot reason about brute-force attackers.
Prior work addresses this gap only through protocol-specific, manually encoded reveal rules, which neither generalize across protocols nor systematically explore compound-key decomposition.
We introduce the first protocol-agnostic, automated symbolic methodology for low-entropy keys,
built on three abstractions: the message-derivation fact \texttt{MDerive}, which captures
brute-force prerequisites independently of concrete function symbols; a three-tier breaking-oracle
design (value-breaking, relationship-breaking, message-derivation) that composes with the standard
Dolev-Yao attacker; and a graded bit-level attacker model that bounds the key length an attacker
is allowed to exhaust.
We realize the methodology as EntropyVerif, a syntax-sugar extension of \Tamarin{} that requires no kernel modification.
Using EntropyVerif we automatically rediscover the PMKID, KNOB, and Method Confusion attacks, and
uncover a new attack on BC which we call the Unilateral Multi-Downgrade (UMD) attack: by
repeatedly downgrading one side of a BC session and recombining the recovered key fragments, an
attacker reduces session-key recovery against devices using the maximum 128-bit key from $2^{128}$
to $n \cdot 2^{128/n}$, bottoming out at $2^{12}$ for $n=16$---a factor of $2^{116}$ reduction
that has been responsibly disclosed to the Bluetooth SIG.
Across three real-world protocols and six BC device configurations, our results show that entropy-aware symbolic analysis is both feasible and necessary.
\end{abstract}

\IEEEpeerreviewmaketitle

\section{Introduction}

Entropy is a fundamental concept that is directly correlated with the security of the cryptographic system.
A key with high entropy indicates a high level of difficulty for an attacker to recover the key from the ciphertext.
However, in practice, modern cryptographic protocols, such as WPA2-PSK, Bluetooth Low Energy~(BLE), and Bluetooth Classic~(BC), usually tolerate low-entropy keys to accommodate user-friendly designs, maintain backward compatibility, or support resource-constrained devices.
This raises a security issue concerning the recovery of low-entropy keys in these protocols, as demonstrated by attacks like PMKID~\cite{pmkidattack} and KNOB~\cite{antonioli2019knob}.

The use of symbolic models to formally analyze protocol vulnerabilities, pioneered by Dolev et al.~\cite{dolev1983security}, has become a mainstream research approach.
Numerous tools have been developed to automatically analyze protocols and identify potential attacks, including AVISPA\cite{armando2005avispa}, ProVerif~\cite{blanchet2018proverif}, Maude-NPA~\cite{escobar2007maude}, Scyther~\cite{cremers2008scyther}, \Tamarin{}~\cite{meier2013tamarin}, and others.
In symbolic models, messages and keys are abstracted as terms in term algebra, rather than as concrete bit sequences as used in computational models.
Consequently, the entropy of keys is lost during this abstraction process, making it impossible for existing symbolic model tools to distinguish low-entropy keys or analyze the brute-force capabilities of attackers.

This raises a critical question: How to capture the brute-force behaviors of attackers within symbolic models?
Some studies\cite{shi2023blesc, claverie2023tamarin} have proposed solutions by manually identifying keys that are vulnerable to compromise and defining specific rules that allow these keys to be directly revealed to attackers.
However, these studies do not provide a general or automated method for analyzing the impact of low-entropy keys on the entire protocol.
Firstly, this method relies on the manual identification of key vulnerabilities, limiting the scalability and objectivity of the analysis.
Secondly, manual analysis is inherently constrained, as it only offers attacks a limited set of predefined ways to compromise low-entropy keys.
As a result, it fails to provide a comprehensive strategy for effectively exploiting such vulnerabilities.

To automatically analyze protocols tolerating low-entropy keys, we encounter three main challenges:
\begin{enumerate}[leftmargin=*]
    \item \textbf{Entropy Representation:}
    Symbolic analysis tools typically preserve only the logical structure of messages or keys, making it impossible to deduce and analyze the entropy of keys.
    Additional mechanisms are needed to describe a key's entropy in order to cover brute-force
    attacks within the symbolic model.
    \item \textbf{Prerequisites for Brute-Force Attacks:}
    Brute-force attacks involve an attacker repeatedly attempting keys, relying on prior knowledge of the algorithm, its outputs, and other relevant inputs beside targeted keys~(e.g., salt values, initialization vectors). 
    Capturing these prerequisite conditions within a symbolic model,
    without tying them to one protocol, presents a significant challenge,
    as these conditions vary across different protocols and keys.
    \item \textbf{Enhanced Adversarial Capabilities:}
    In existing works, an attacker is typically limited to compromising specific keys. 
    However, an attacker can attempt to brute-force any key as long as its entropy doesn't exceed its computational capacity. 
    The attacker can also deconstruct complex keys into simpler sub-keys and brute-force them individually. 
    A challenge lies in modeling this flexible brute-forcing capability within the symbolic model.
\end{enumerate}

This paper aims to bridge the existing gap by introducing a protocol-independent and automated
method for modeling low-entropy keys and the adversarial brute-force capabilities within symbolic
analysis frameworks.
We focus on enhancing \Tamarin{} to describe brute-force attacks without encoding them separately
for each protocol.
\Tamarin{} is a state-of-the-art protocol verification tool widely used for the formal analysis of many protocols, including TLS\cite{cremers2017tls13}, 5G-AKA\cite{basin20185G, wang20215G}, EMV\cite{basin2021emv,radu2022emv}, etc.
We have designed a new syntax, which is a superset of the syntax of \Tamarin{}, allowing for the labeling of low-entropy variables in symbolic models. 
Additionally, we have developed an automated protocol analysis tool supporting such low-entropy annotated models, which we call EntropyVerif\cite{toolrepo}. 
This tool supports an adversary model that extends beyond the Dolev-Yao model,
capable of capturing brute-force behaviors that a deterministic attacker can carry out by
recomputing a function forward and comparing the result against an observed output,
including attacks in which the attacker combines several such recoveries across sessions.
Our approach can be used to identify security issues arising from low-entropy keys in protocols
and to check protocols that tolerate low-entropy keys for attacks within a stated attacker model
and budget.

\smallskip
\noindent\textbf{A new attack on Bluetooth Classic.}
Applying EntropyVerif to BC session establishment immediately exposes a new protocol-level attack that we call the Unilateral Multi-Downgrade (UMD) attack.
The BC specification allows the two endpoints to negotiate a session-key length between 1 and 16 bytes.
UMD exploits the fact that when one side is repeatedly renegotiated down to a short key, the resulting session keys are not independent fresh randomness but correlated fragments derived from a shared long-term secret; an attacker can therefore brute-force each fragment separately and stitch the full-length session key back together.
To our knowledge this is the first protocol-level attack on BC discovered by a symbolic verifier; Section~\ref{sec:umd-attack} gives the full complexity analysis and the attack trace.
The attack has been responsibly disclosed to the Bluetooth SIG.

\noindent\textbf{Contributions.}
Our primary contributions are as follows:
\begin{itemize}[leftmargin=*]
    \item \emph{Protocol-agnostic methodology.} We relax the \emph{perfect encryption assumption} in the symbolic model and introduce the \texttt{MDerive} message-derivation fact together with a three-tier breaking-oracle design (value-breaking, relationship-breaking, message-derivation). Unlike prior ad-hoc extensions~\cite{shi2023blesc,claverie2023tamarin,jangid2023bluepe} that hard-code reveal rules for a specific protocol, our abstractions are function-symbol agnostic and support recursive decomposition of compound keys.
    \item \emph{Tool without kernel modification.} We developed EntropyVerif, a syntax-sugar extension of \Tamarin{}, in which users annotate low-entropy terms with \texttt{:= L}.
          \emph{The compiler emits standard multiset-rewriting rules and leaves the \Tamarin{}
          backend unmodified, so every proof is carried out by \Tamarin{} itself.}
    \item \emph{Rediscovery of known attacks.} Using EntropyVerif we automatically reproduce the PMKID attack on WPA2-PSK, the KNOB attack and the Method Confusion attack on BLE pairing, without protocol-specific reveal rules.
    \item \emph{A new attack on Bluetooth Classic (UMD).} Beyond rediscovery, the compound-key decomposition naturally supported by \texttt{MDerive} leads the verifier to a previously unknown protocol-level attack: the Unilateral Multi-Downgrade attack on BC, which reduces 128-bit session-key recovery from $2^{128}$ to $n \cdot 2^{128/n}$, i.e.\ as low as $2^{12}$ when $n = 16$---a factor of $2^{116}$ reduction. The attack has been responsibly disclosed to the Bluetooth SIG.
\end{itemize}

\section{Background} \label{sec:2background}

\subsection{Entropy and Brute-Force Attacks}

Entropy was introduced by Shannon~\cite{shannon1948communicationtheory} to quantify the randomness and uncertainty within a system.
In the domain of cryptography, the entropy of a key serves as a crucial metric for assessing its strength.

The method of enumerating all possible keys until the correct one is found is called a brute-force attack, which can be classified as either an online or an offline attack.
In an online attack, the adversary persistently sends the attempted keys to the target system to verify their correctness.
In an offline attack, the attacker need only obtain the ciphertext and test all possible keys to decrypt it.
In this paper, we focus on offline brute-force attacks; all subsequent references to brute-force attacks refer to the offline variant unless stated otherwise.

\subsection{Protocols Tolerating Low-Entropy Keys}
Low-entropy keys are common in practice.
Backward compatibility, power consumption constraints, and user experience requirements prevent many protocols from guaranteeing high-entropy keys for all cryptographic variables.
We examine three representative protocols: WPA2-PSK, Bluetooth Low Energy (BLE), and Bluetooth Classic (BC).

\subsubsection{WPA2-PSK}
\begin{figure}[t]
    \centering
    \includegraphics[width=0.870780\columnwidth]{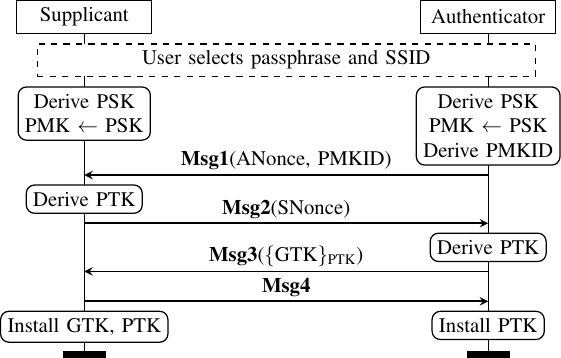}
    \caption{Overview of WPA2-PSK}
    \label{fig:4-way-handshake}
\end{figure}

Wi-Fi Protected Access 2 Pre-Shared-Key~(WPA2-PSK) establishes a secure connection between a supplicant and an authenticator in a personal Wi-Fi network.
\autoref{fig:4-way-handshake} illustrates the protocol, where both parties derive a Pairwise Transient Key~$PTK$ from a Pairwise Master Key~$PMK$, which is computed from a pre-shared $Passphrase$ and the network name~$SSID$:
\begin{equation} \label{eq:pmk}
    PMK = \text{PBKDF2}(Passphrase, SSID)
\end{equation}

The $Passphrase$ consists of 8--63 characters; a user who chooses digits only significantly reduces the entropy of $PMK$.
In the 4-way handshake, the authenticator and supplicant exchange nonces $N_A, N_B$ and derive $PTK = \text{PRF}(PMK, {MAC}_A, {MAC}_S, N_A, N_B)$.
The first EAPOL-key message carries a $PMKID$ field for fast re-authentication, computed as:
\begin{equation} \label{eq:pmkid}
    PMKID = \text{HMAC}(PMK, {MAC}_{A} \| {MAC}_{S})
\end{equation}
Messages~2--4 are protected by a MIC derived from $PTK$, but the $PMKID$ in message~1 is sent before $PTK$ exists and thus lacks integrity protection.
This field is exploited by the PMKID attack~(\S\ref{sec:wpa2-psk}).

\subsubsection{Bluetooth Low Energy~(BLE)}
\begin{figure}[t]
    \centering
    \includegraphics[width=0.912689\columnwidth]{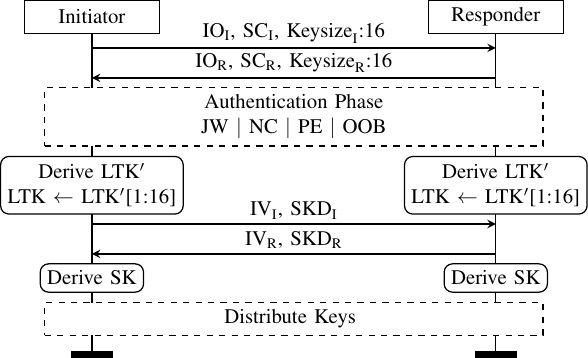}
    \caption{Overview of BLE pairing protocol}
    \label{fig:ble-pairing}
\end{figure}

BLE~\cite{bluetooth4} is designed for short-range, low-power communication.
The BLE pairing protocol negotiates a long-term key~$LTK$ between an initiator and a responder; $LTK$ is then used to derive a session key~$SK$ for an encrypted channel.
\autoref{fig:ble-pairing} shows the pairing process.

During feature exchange, devices announce their supported key sizes (7--16 bytes) and IO capabilities.
An intermediate key $LTK'$ is negotiated during authentication, then truncated to the agreed $KeySize$ to form $LTK$.
Both devices then compute $SK = \text{AES-ECB}_{LTK}(SKD_I \| SKD_R)$, where $SKD_I, SKD_R$ are random contributions.
The key-size negotiation and truncation phases are unprotected by integrity checks, enabling an attacker to force a shorter key.

\subsubsection{Bluetooth Classic~(BC)}
\begin{figure}[t]
    \centering
    \includegraphics[width=0.902212\columnwidth]{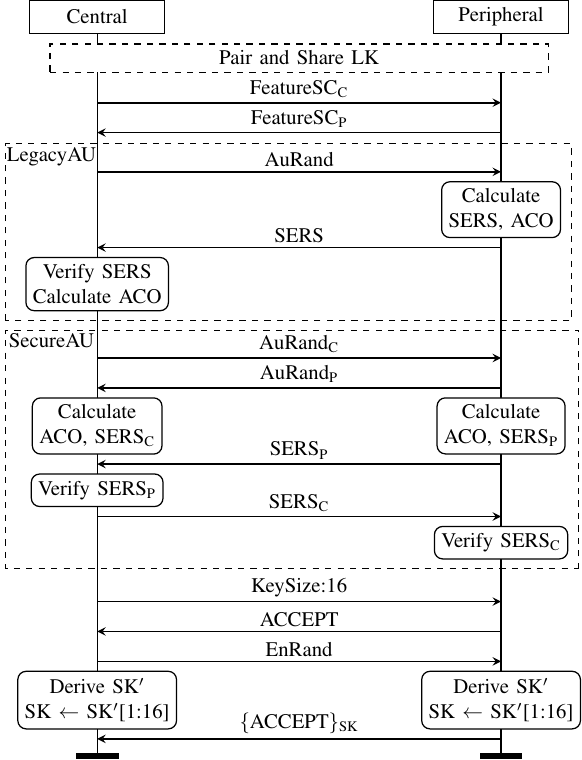}
    \caption{Overview of BC session establishment protocol}
    \label{fig:bc-session}
\end{figure}

Bluetooth Classic (BR/EDR) is designed for high-throughput applications such as audio streaming and file transfer.
\autoref{fig:bc-session} shows the session establishment between paired BC devices.

After pairing, both devices share a link key~$LK$.
Session establishment proceeds through four phases: (1)~link connection, where devices exchange supported features including whether they support \emph{Secure Connections}; (2)~authentication, using either Legacy Authentication (unilateral) or Secure Authentication (mutual) to compute an Authenticated Channel Offset~$ACO$ from $LK$ and random values; (3)~key-size negotiation; and (4)~session-key derivation, where the devices compute a session key $SK'$ from $LK$ and $ACO$:
\begin{equation} \label{eq:legacy-sk}
    \text{Legacy:}\; SK' = \text{E3}(LK, EnRand, ACO)
\end{equation}
\begin{equation} \label{eq:secure-sk}
    \text{Secure:}\; SK' = \text{h3}(LK, ACO)
\end{equation}
The final session key~$SK$ is obtained by truncating $SK'$ to the negotiated key length.
The key-size negotiation phase is unprotected, and the encrypted \Tfunction{ACCEPT} message using $SK$ concludes the session establishment.

\subsection{\Tamarin{}}
\Tamarin{}~\cite{meier2013tamarin} is a state-of-the-art tool for formal security protocol verification.
Protocols and adversary behaviors are modeled as labeled multiset rewriting rules of the form $[F_l] \xrightarrow{[F_a]} [F_r]$, where $F_l, F_a, F_r$ are multisets of facts.
A \emph{fact} $F(t_1, \ldots, t_n)$ is built from terms in a term algebra $T_\Sigma(\mathcal{V}, \mathcal{N})$ over function symbols~$\Sigma$, variables~$\mathcal{V}$, and names~$\mathcal{N}$.
Facts are either \emph{linear} (consumed once) or \emph{persistent} (reusable).
\Tamarin{} provides built-in facts \TFact{In} (receive from public channel), \TFact{Out} (send), and \TFact{Fr} (fresh name generation).
Security properties are expressed as first-order trace formulas with time points; for instance, secrecy of a key~$k$ requires that no adversary learns $k$ after it is installed:
\begin{flalign*}
    & \hspace{0em} \TAll \;\, \text{\Tterm{k \#i.}} \;\, \text{\TFact{Install(k)@\#i}} &\\
    & \hspace{3em} \implies \TNot (\TEx \;\, \text{\TLemma{\#j.}} \;\, \text{\TFact{K(k)@\#j}} ) &
\end{flalign*}

\section{Brute-Force Attack Modeling} \label{sec:3symbolic-modeling}

In this section, we will discuss the limitations of existing models and introduce our methodology
for capturing brute force behaviors without encoding them separately for each protocol.

\subsection{Threat Model and Low-Entropy Adversary} \label{sec:threat-model}

Our adversary extends the classical Dolev-Yao attacker~\cite{dolev1983security}
with a bounded brute-force capability.
As in the Dolev-Yao model, the attacker fully controls the public network: they can intercept, drop, reorder, forge, and replay messages, and they can derive any term that is the result of applying a public function to knowledge already in their possession.
Beyond these capabilities, the attacker may attempt to recover the value of a term $k$ from a function output $f(\ldots, k, \ldots)$ whenever (i)~$k$ has been labeled as low-entropy in the protocol model, and (ii)~every non-$k$ argument of $f$ is either a constant or derivable by the Dolev-Yao attacker.
The recovery is deterministic and forward: for each candidate value of $k$ the attacker recomputes
$f$ and keeps the candidate whose output matches the observed one exactly.
It never applies an inverse of $f$ to the observed output, and checks a candidate by recomputing
$f$ rather than by inverting it.
We call an attack that reaches its goal through one or more such recoveries a \emph{forward
brute-force attack}, and this is the class of attacks we capture.

Such a recovery takes up to $2^{|k|}$ trials, where $|k|$ is the nominal bit-length of $k$,
and is admitted only if $|k|$ lies within a budget fixed in advance by the modeler.
To anchor that budget at a concrete bit level we introduce two graded adversaries used throughout the case studies.
A \emph{semi-attacker} (SA) is budgeted to $\lambda_{\mathrm{SA}} = 64$ bits and can therefore brute-force any key up to 8 bytes in length, matching the computational scale of the documented KNOB attack.
A \emph{quart-attacker} (QA) is budgeted to $\lambda_{\mathrm{QA}} = 32$ bits and can brute-force any key up to 4 bytes, representing a lower-resource adversary.
Both adversaries retain the full Dolev-Yao repertoire in addition to their brute-force oracle; the difference is purely the maximum $|k|$ at which the oracle is allowed to fire.
Sections~\ref{sec:4extension-and-compilation} and~\ref{sec:5case-study} formalize this behavior as multiset-rewriting rules and instantiate SA/QA against concrete BC device configurations.

\subsection{Key-Breaking Oracles}

\emph{Perfect encryption assumption} adopted by current symbolic models prevent adversaries from obtaining any information from the ciphertext.
\cite{cremers2023aead,shi2023blesc,claverie2023tamarin} proposed their methods to loosen this assumption by introducing a set of special rules.
These rules model the behaviors of attackers accessing an oracle, thereby allowing them to obtain certain keys.
Such a rule of \cite{claverie2023tamarin} is as follows:
\begin{flalign*}
    \text{\ruleNameText{Rul}} & \text{\ruleNameText{e Oracle}} :                    &\\
    \TFacts{ &\text{\TFact{LowEntropyf4}}(pk1, pk2, n, s), \\
    &\text{\TFact{In}}(pk1), \text{\TFact{In}}(pk2), \text{\TFact{In}}(n), \text{\TFact{In}}(f4(pk1, pk2, n, s))} \TEmpActFact{ } \\
    \TFacts{ &\text{\TFact{Out}}(s)}    
\end{flalign*}
In this rule, the fact \TFact{LowEntropyf4} indicates that the last input $s$ of the function $f4$ possesses a low entropy.
Attackers can recover the value of $s$ by providing all other inputs of $f4$ and its output.
This rule is intuitive and checks the three essential prerequisites for a brute-force attack: knowledge of the function, knowledge of all inputs except the key, and knowledge of the computation result. 
It effectively captures brute-force attacks targeting the $f4$ function.

However, this is not a general modeling method.
The design of this rule implicitly relies on a critical hypothesis: an attacker is capable of, and restricted to, exploiting the function $f4$ to execute a brute force attack on the low-entropy variable $s$.
This hypothesis is confirmed manually, rather than automatically.
This presents a challenge, as the inability to ascertain which function an attacker might utilize to brute force the key renders the application of such a rule infeasible.

To capture forward brute-force attacks, we designed a key-breaking oracle.
This oracle accepts any function symbol declared in the model and returns the value of the
low-entropy key, provided that the attacker can furnish the function output along with the values
of all other arguments, as shown below:
\begin{figure}[h]
    \centering
    \includegraphics[width=0.7755\columnwidth]{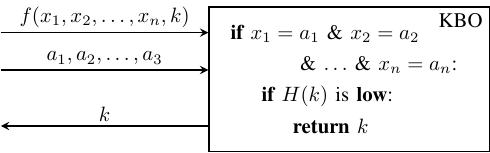} 
\end{figure}

\noindent
To model this oracle, there are two challenges.
\begin{enumerate}[leftmargin=*]
    \item \textbf{Uncertainty of Received Functions:} The oracle is not written for one protocol,
    so it has to accept any function symbol the model declares.
    But in symbolic models, function symbol matching is precise, prohibiting using a single function symbol to generally match all functions.
    \item \textbf{Uncertainty of Additional Inputs:} Given the uncertainty regarding which function will be received, the oracle must dynamically accommodate additional inputs contingent upon the specific function it encounters.
\end{enumerate}
To address the first challenge, we introduce the concept of message derivation relationships as a substitute for specific function symbols. 
This concept will be explained in detail in \autoref{sec:oracle-modeling}.
To overcome the second, we introduce two more fundamental oracles: the value-breaking oracle and the relationship-breaking oracle. 

\subsubsection{Value-Breaking Oracle}
The execution process of the value-breaking oracle is delineated as follows:
\begin{figure}[h]
    \centering
    \includegraphics[width=0.4653\columnwidth]{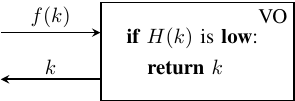} 
\end{figure}

\noindent The value-breaking oracle grants attackers the capability to break low-entropy keys, assuming the attacker can identify a computation result of a function in which a low-entropy key is the only input. 
An attacker submits to the value-breaking oracle the output of such a function output.
The only thing the value-breaking oracle needs to do is to verify whether the key possesses a low entropy and return the key's value if it does.

\subsubsection{Relationship-Breaking Oracle}
The relationship-breaking oracle is designed to eliminate the attacker's known variables from the key's computation.
The execution process of the relationship-breaking oracle is shown as follows:
\begin{figure}[h]
    \centering
    \includegraphics[width=0.99\columnwidth]{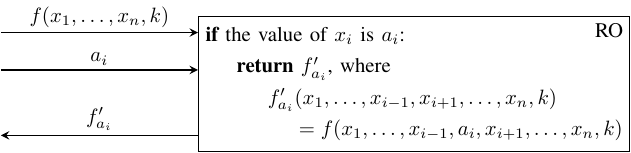} 
\end{figure}

\noindent
Consider a function $f$ of arity $n+1$ whose output $f(x_1,\ldots,x_n,k)$ the attacker has
observed, where $k$ is the target key and $x_1,\ldots,x_n$ are the remaining arguments.
Suppose the attacker learns that argument $x_i$ has the value $a_i$ and submits that value
together with the observed output to the relationship-breaking oracle.
The oracle returns the partial application of $f$ in which the $i$-th argument is fixed to $a_i$:
\begin{flalign*}
    & f'_{a_i}(x_1,\ldots,x_{i-1},x_{i+1},\ldots,x_n,k) &\\
    & \quad = f(x_1,\ldots,x_{i-1},a_i,x_{i+1},\ldots,x_n,k). &
\end{flalign*}
The subscript is part of the definition. Fixing the same position to a different value yields a
different function, so $f'$ is determined by $f$, the position $i$, and the value $a_i$ together.
Repeating this removes known arguments one at a time until $k$ is the only argument left, at which
point the attacker calls the value-breaking oracle on the remaining output.

\subsection{Oracle Modeling} \label{sec:oracle-modeling}

We model the previous oracles through a series of rules.
To abstract the process of message derivation within the relationship-breaking oracle, we introduce the concept of message derivation facts, replacing specific function symbols.
Furthermore, to model the relationship-breaking oracle, it is necessary to incorporate message derivation rules that enable the oracle to deduce all message derivation facts visible from the attacker's perspective.
Building on this foundation, we will present our approach to modeling the value-breaking oracle.

\subsubsection{Message Derivation Fact} In the formalization of the relationship-breaking oracle, key computation is abstracted into derivation relationships, represented by message derivation facts.
Each derivation fact maps one message to another.
Specifically, for a one-arity function $f$ and an argument $x$, $f$ can be abstracted as the fact $\text{\TFact{MDerive}}(x, f(x))$.
For the sake of consistency, multi-arity functions are modeled as:
\[
    \text{\TFact{MDerive}}(\langle x_1, x_2, ..., x_n \rangle, f(\langle x_1, x_2, ..., x_n \rangle)),
\]
where $\langle \cdot \rangle$ denotes a function of $n$ arguments, representing the concatenation of $n$ function arguments.
Similarly, a composite function $f \circ g$ can be modeled as the message derivation fact $\text{\TFact{MDerive}}(x, f(g(x)))$.

The relationship-breaking oracle maintains a collection of all function symbols declared in the
model.
Once a derivation relationship between two messages is identified, the oracle can locate the specific mapping function.
If the oracle needs to identify a function where a certain key $k$ is the sole argument, it suffices to find the fact $\text{\TFact{MDerive}}(k, m)$,
where $m$ is a message derived from $k$.

\subsubsection{Message Derivation Rule}  
Upon the initiation of the attacker's access to the oracle, the oracle acquires all derivation facts present in the current execution process of the protocol.
It then infers all existing derivation facts that are computable from the attacker's perspective.

Formally, given a set $ df_\mathcal{O} $ denotes all message derivation facts known to the oracle, while $ dm_\mathcal{A} $ represents the set of all deducible messages that the adversary can know.
When the adversary $ \mathcal{A} $ initiates interaction with the oracle $ \mathcal{O} $, the oracle gathers all message derivation facts from the current execution of the protocol $\mathcal{P}$, denoted by $df_\mathcal{P}$, and uses these to initialize $ df_\mathcal{O} $.
The set $ df_\mathcal{O} $ is inductively defined by the following rules:
\begin{align*}
    &\infer
        {\text{\TFact{MDer}}(m_1, m_2) \in df_\mathcal{O}}
        {\text{\TFact{MDer}}(m_1, m_2) \in df_\mathcal{P}}
    &&
    \infer
        {\text{\TFact{MDer}}(m_2, m_3) \in df_\mathcal{O}}
        {\begin{array}{c} 
            m_1 \in dm_\mathcal{A} \\ 
            \text{\TFact{MDer}}(\langle m_1, m_2 \rangle, m_3) \in df_\mathcal{O}
        \end{array}}
\end{align*}
\begin{align*}
    &\infer
        {\text{\TFact{MDer}}(m_1, m_3) \in df_\mathcal{O}}
        {\begin{array}{c} 
            \text{\TFact{MDer}}(m_1, m_2) \in df_\mathcal{O} \\ 
            \text{\TFact{MDer}}(m_2, m_3) \in df_\mathcal{O}
        \end{array}} 
    &&
    \infer
        {\text{\TFact{MDer}}(m_1, m_3) \in df_\mathcal{O}}
        {\begin{array}{c} 
            m_2 \in dm_\mathcal{A} \\ 
            \text{\TFact{MDer}}(\langle m_1, m_2 \rangle, m_3) \in df_\mathcal{O}
        \end{array}}
\end{align*}

\noindent
In these rules, the attacker is required to continuously supply all known terms to assist the oracle in inferring all visible derivation facts.

\subsubsection{Value-Breaking Oracle} 
In the concluding step of the oracle's key-breaking process, the oracle verifies whether the target key possesses a low entropy and whether the attacker has knowledge of a calculation result where the key is the only variable.
If these prerequisites are satisfied, the oracle will return the value of the key.
This rule can be stated as follows:
\begin{flalign*}
    \text{\ruleNameText{Rul}} & \text{\ruleNameText{e OracleKeyCrack}} :                               &\\
    \TFacts{ &\text{\TFact{MDerive}}(k, m), \text{\TFact{LowEntropy}}(k), \text{\TFact{In}}(m) } \TEmpActFact{ }  \\
    \TFacts{ &\text{\TFact{Out}}(k) }        
\end{flalign*}
where $k$ denotes the attacker's target key, $m$ represents the computation result from the key, and $\text{\TFact{LowEntropy}}(k)$ labels that $k$ is a low-entropy key, which is specified during modeling a certain protocol.

\section{Low-Entropy Analysis Methodology} \label{sec:4extension-and-compilation}
In this section we describe how to apply our oracle modeling approach to analyze low-entropy keys in protocols.
To enhance the modeling of protocols using our methodology, we have devised an innovative rule syntax that serves as a superset of \Tamarin{}'s syntax. 
This new syntax incorporates several new keywords designed to identify low-entropy keys that require special attention. 
Additionally, to facilitate our new syntax, we have developed an automated low-entropy analysis tool, EntropyVerif. 
This tool integrates the key-breaking oracle discussed in the previous section, enabling attackers
to compromise low-entropy keys, thus providing a security analysis of protocols tolerant to such
keys.
To support the operation of the key-breaking oracle, our tool automatically analyzes the derivation relationships among variables within the model and generates message derivation facts. 
Finally, to ensure the automation and efficiency of the verification process, we have introduced restrictions on the oracle rules and explained these restrictions.

\subsection{ Labeled Rule Syntax } \label{sec:syntax-extension}

\Tamarin{} utilizes multiset rewriting rules to define protocol execution and adversarial behaviors.
For instance, consider the following rule demo:
\begin{flalign*}
\text{\Tterm{Rule S}}  & \text{\Tterm{endHello}} :                                      & \\
    \TLetS & \text{\Tterm{$ciphertext$ = senc("hello", $key$)}}   \\  
    \TInS & \TFacts{ \text{\TFactTerm{Share}{$key$}} }  \TEmpActFact{} \TFacts{ \text{\TFactTerm{Out}{$ciphertext$}} }
\end{flalign*}
This rule delineates a simple behavior for sending encrypted data: an entity encrypts the message "hello" using a shared $key$ and sends the $ciphertext$ to a public network.
\Tamarin{} supports local macro definitions in a rule which are defined within the \TFact{let}$ \ldots $\TFact{in} area called the let-block, where each line specifies a substitution.
In the above demo, \Tterm{$ciphertext$} in this rule will be replaced by \Tterm{senc("hello", $key$)} when the rule is executed.

To apply our methodology, we introduced a new operator ``:='' and a keyword ``\verb|`L`|''.
These syntax elements facilitate the identification and labeling of low-entropy variables.
In the extended syntax, as depicted in the following rule, the low-entropy label for the variable is defined within the let block.
\begin{flalign*}
    \text{\Tterm{Rule S}}   & \text{\Tterm{endHello}} :                         & \\
    \TLetS                  & \text{\Tterm{$key$ := `L`}}                         \\  
                            & \text{\Tterm{$ciphertext$ = senc("hello", $key$)}}    \\  
    \TInS                   & \TFacts{ \text{\TFactTerm{Share}{$key$}} }  \TEmpActFact{} \TFacts{ \text{ \TFactTerm{Out}{$ciphertext$}} }
\end{flalign*}
Here, the label ``\verb|`L`|'' on the right side of ``:='' signifies that the variable $key$ on the left possesses a low entropy.
This label will be recognized by EntropyVerif and translated into a fact $\text{\TFact{EntropyLow}}(key)$ added to the right-hand side of the rule.

\subsection{ Design of EntropyVerif }

\begin{figure}[t]
    \centering
    \includegraphics[width=\columnwidth]{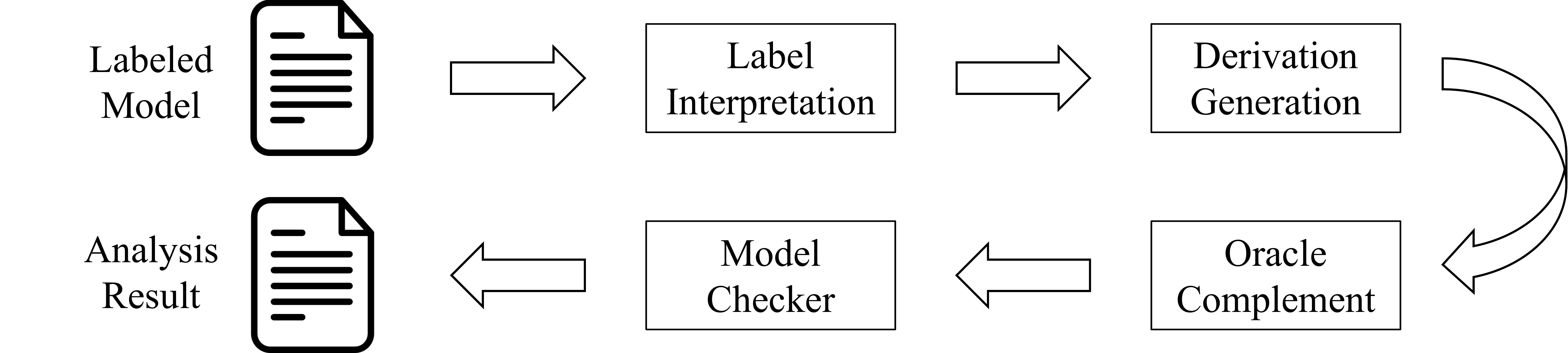}
    \caption{Architecture of EntropyVerif}
    \label{fig:design-of-compilation-tool}
\end{figure}

EntropyVerif is an extension of \Tamarin{} and 
supports low-entropy labeled protocol models and integrates an adversarial model with the capability to brute-force low-entropy keys.
As shown in \autoref{fig:design-of-compilation-tool}, EntropyVerif contains 4 modules:
\begin{enumerate}[leftmargin=*]
    \item \textbf{Label Interpretation:} This module is responsible for identifying and parsing the low-entropy labels of terms, subsequently generating the \TFact{EntropyLow} facts.
    \item \textbf{Derivation Generation:} This module analyzes the message derivation relationships between terms in symbolic models and generates \TFact{MDerive} facts.
    \item \textbf{Oracle Complement:} This module adds the key-breaking oracle to the model and incorporates reasonable restrictions to prevent automated verification from entering a non-terminating state.
    \item \textbf{Verifier:} This module conducts a formal analysis of the model generated by preceding modules. 
\end{enumerate}

We successfully implemented Label Interpretation, Derivation Generation, and Oracle Complement modules, while employing \Tamarin{} as the Verifier in our study.

\subsection{ Analysis Details }

\begin{table}[t]
    \centering
    \caption{Notation used throughout the methodology.}
    \label{tab:notation}
    \renewcommand{\arraystretch}{1.15}
    \small
    \begin{tabular}{@{}ll@{}}
    \toprule
    Symbol & Meaning \\
    \midrule
    $\mathsf{sub}(t)$           & Syntactic subterms of $t$ \\
    $\mathsf{atoms}(t)$         & Atomic terms occurring in $t$ \\
    $\mathsf{vars}(t)$          & Atomic terms of $t$ that are not constants \\
    $\mathsf{const}(T)$         & Constants and public names in $T$ \\
    $\mathsf{terms}(F)$         & Terms occurring as arguments of the facts in $F$ \\
    $df_\mathcal{P}$            & Derivation facts emitted from protocol rules \\
    $df_\mathcal{O}$            & Oracle-side closure of derivation facts \\
    $dm_\mathcal{A}$            & Messages derivable by the Dolev--Yao attacker \\
    \bottomrule
    \end{tabular}
\end{table}

To describe the translation we use the operations listed in Table~\ref{tab:notation}, all of which
are defined on sets.
Throughout, a term is \emph{atomic} if it is a variable, a fresh name, or a public constant. Every
other term has the form $f(t_1,\ldots,t_n)$ for a function symbol $f$ declared in the model.

For a term $t$, $\mathsf{sub}(t)$ is the set of syntactic subterms of $t$:
\[
    \mathsf{sub}(t) =
    \begin{cases}
        \{t\} & \text{$t$ atomic,} \\[2pt]
        \{t\} \cup \bigcup_{i=1}^{n} \mathsf{sub}(t_i) & \text{$t = f(t_1,\ldots,t_n)$.}
    \end{cases}
\]
The definition is purely syntactic. It does not close the set under the equational theory, and it
does not refer to what an attacker is able to derive. Both of those are left to \Tamarin{} during
verification.

Dually, $\mathsf{atoms}(t)$ is the set of atomic terms out of which $t$ is built:
\[
    \mathsf{atoms}(t) =
    \begin{cases}
        \{t\} & \text{$t$ atomic,} \\[2pt]
        \bigcup_{i=1}^{n} \mathsf{atoms}(t_i) & \text{$t = f(t_1,\ldots,t_n)$.}
    \end{cases}
\]
Reading $t$ as a tree, $\mathsf{sub}(t)$ collects every node and $\mathsf{atoms}(t)$ collects the
leaves. Both operations are lifted to a set of terms $T$ by taking the union over its elements, so
that $\mathsf{sub}(T) = \bigcup_{t \in T} \mathsf{sub}(t)$, and likewise for $\mathsf{atoms}$.

We write $\mathsf{const}(T)$ for the subset of $T$ consisting of constants and public names, and
abbreviate $\mathsf{vars}(t) = \mathsf{atoms}(t) \setminus \mathsf{const}(\mathsf{atoms}(t))$ for
the non-constant leaves of $t$. For a collection of facts $F$, $\mathsf{terms}(F)$ is the set of
terms occurring as arguments of those facts.

As in \Tamarin{}, the two sides of a multiset rewriting rule are multisets of facts, written
$[F_L] \rightarrow [F_R]$, and $\uplus$ denotes multiset union. Every other collection in this
section is a set. A fact argument must be a single term, so wherever a set of terms has to be
passed to a fact we arrange it in a fixed order and pass the resulting tuple, written $\langle
\cdot \rangle$. That order belongs to the encoding and carries no meaning.

Based on the aforementioned notations, we explain how EntropyVerif verifies protocol models with low-entropy labels.

\noindent
\textbf{Label Interpretation:} 
Labels of low-entropy terms are not inherently part of the multiset rewriting system. 
It is introduced to facilitate the identification of such terms within existing rules.
For a given rule $[F_L] \rightarrow [F_R]$, $T_{Low}$ denotes the set of low-entropy terms labeled
in the rule, where $T_{Low} \subseteq \mathsf{sub}(\mathsf{terms}(F_L)) \cup
\mathsf{sub}(\mathsf{terms}(F_R))$.
The translated version of such a rule with low-entropy labels is shown as follows:
\[
    [F_L] \rightarrow [F_R \uplus \text{\textsf{LE}}(T_{Low})]
\]
where, for brevity, we write $\text{\textsf{LE}}(T_{Low})$ for the multiset $\{\,
\text{\TFact{EntropyLow}}(t) \mid t \in T_{Low} \,\}$.

We take the rule labeled with low entropy in \autoref{sec:syntax-extension} as an example. 
In the Label Interpretation, it will be translated into the following rule:
\begin{flalign*}
    \text{\Tterm{Rule S}}   & \text{\Tterm{endHello}} :                         & \\
    \TLetS                  & \text{\Tterm{$ciphertext$ = senc("hello", $key$)}}    \\  
    \TInS                   & \TFacts{ \text{\TFactTerm{Share}{$key$}} }          \\
    \TEmpActFact{} \hspace{0.2em}& \TFacts{ \text{ \TFactTerm{Out}{$ciphertext$} }, \text{ \TFactTerm{EntropyLow}{$key$} }  }
\end{flalign*}

\noindent
\textbf{Derivation Generation:} 
Given the rule $[F_L] \rightarrow [F_R]$, $\mathsf{terms}(F_R)$ denotes the terms present within
the rule after executing, whereas $\mathsf{sub}(\mathsf{terms}(F_L))$ represents all possibly
accessible terms before the execution.

Thus, $\mathsf{terms}(F_R) \setminus \mathsf{sub}(\mathsf{terms}(F_L))$ can be used to represent
the new terms derived during the execution of the rule.
We define the rules incorporating message derivation facts as follows:
\[
    [F_L] \rightarrow [F_R \uplus \text{\textsf{MD}}(\mathsf{terms}(F_R) \setminus \mathsf{sub}(\mathsf{terms}(F_L)))]
\]
Here, $\text{\textsf{MD}}(T)$ denotes the set of message derivation facts.
For an individual term $t$, $\mathsf{vars}(t)$ contains the non-constant leaves needed to
construct it.
Hence $\text{\textsf{MD}}(T)$ can be defined as
$\{\, \text{\TFact{MDerive}}(\langle \mathsf{vars}(t) \rangle, t) \mid t \in T \,\}$.

We take the rule in \autoref{sec:syntax-extension} as an example as well. 
A message derivation fact will be added to the right-hand side of the rule as follows:
\begin{flalign*}
    \text{\Tterm{Rule S}}   & \text{\Tterm{endHello}} :                         & \\
    \TLetS                  & \text{\Tterm{$ciphertext$ = senc("hello", $key$)}}    \\  
    \TInS                   & \TFacts{ \text{\TFactTerm{Share}{$key$}} }          \\
    \TEmpActFact{} \hspace{0.2em}& \TFacts{ \text{ \TFactTerm{Out}{$ciphertext$} }, \text{ \TFactTerm{EntropyLow}{$key$} }, \\
                            & \hspace{5em} \text{ \TFactTerm{MDerive}{$key$, $ciphertext$} }}
\end{flalign*}

\noindent
\textbf{Oracle Complement:} 
The Oracle Complement module adds the key-breaking oracle to protocol models and generates appropriate restrictions.
Despite the aforementioned work enabling the oracle to operate, additional restrictions on the oracle's rules are necessary to ensure automation, and efficiency, and to prevent the system from entering a non-terminating state.

For a given term $t$, $\mathsf{vars}(t)$ contains all variables required to construct it.
Since $\mathsf{vars}(t)$ is finite and each execution of an oracle rule consumes one of its
elements, if $t$ can be compromised by the oracle then the oracle rule is executed at most
$|\mathsf{vars}(t)|$ times.
So in the Oracle Complement module, we reserved a verification option to dynamically restrict the number of times the oracle rule can be executed.

\noindent
\textbf{Verifier:} 
In EntropyVerif, \Tamarin{} is employed as the Verifier to analyze the model generated by the aforementioned modules.
The last module, Oracle Complement, outputs the generated model to the Verifier module in a format compatible with \Tamarin{}. 
The Verifier verifies whether the security properties in the protocol are satisfied and outputs the results to an analysis results file.

\section{ Case Studies } \label{sec:5case-study}

In this section, to demonstrate the effectiveness of our methodology, we will conduct case studies of several real-world protocols that have issues with low-entropy keys. 
In our case studies, we consider an adversary model that builds upon the classical Dolev-Yao model. 
In addition to the standard abilities of a Dolev-Yao attacker, our adversary is equipped with the
capability to execute brute-force attacks on low-entropy keys.
The source code of our tool and models can be found in our repository~\cite{toolrepo}.

\subsection{WPA2-PSK} \label{sec:wpa2-psk}

\begin{figure}[t]
    \centering
    \includegraphics[width=0.925495\columnwidth]{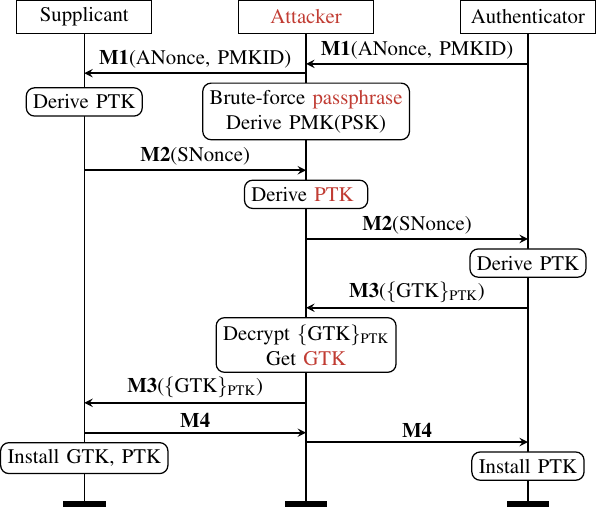} 
    \caption{Attack flow of PMKID attack on WPA2-PSK}
    \label{fig:pmkid-attack}
\end{figure}

The security of WPA2-PSK is protected by the $PMK$. 
The first message of the 4-way-handshake includes $PMKID$ which is calculated from the user chosen $Passphrase$, see~\autoref{eq:pmkid}. 
However, if the user chooses a simple $Passphrase$, the entropy of the PMK will be reduced, allowing an attacker to brute-force $PMK$ from $PMKID$. 
In this case, the attacker can compromise the $PTK$ and $GTK$ through the $PMK$, thereby compromising the security of the entire session. 
This attack is known as the PMKID attack\cite{pmkidattack} and is shown in \autoref{fig:pmkid-attack}.

\noindent
\textbf{Security Assumption:} 
We assume the existence of a security-conscious user who always configures weak $Passphrase$ on the authenticator and the supplicant.

\noindent
\textbf{Protocol Modeling:}  
\cite{cremers2020wpa2} proposed a formal model of the WPA2, including the process of distributing the $PTK$ and $GTK$ with $PMK$.
As the model primarily focuses on verifying key reinstallation attacks, the $PMKID$ is omitted in the first handshake message.
We added $PMKID$ to the first message of the handshake as follows:
\begin{flalign*}
    \text{\ruleNameText{Rule }}    & \text{\ruleNameText{AuthSndM1}} :                                          & \\
    \TLetS                          & \text{\Tterm{$PMK$ := `L`}}                                                 \\  
                                    & \text{\Tterm{$PMKID$ = HMAC("PMK Name", $PMK$)}}                              \\  
    \TInS \TFacts{                  & \text{\TFactTerm{AuthState}{"INIT", $PMK$}}, \; \text{\TFactTerm{Fr}{$ANonce$}}}  \\
    \TEmpActFact{} \hspace{0.2em} \TFacts{& \text{\TFactTerm{AuthState}{"START", \TPairN{$PMK$, $ANonce$}}},            \\
                                    & \text{\TFactTerm{AuthOut}{\TPairN{$ANonce$, $PMKID$}}}}
\end{flalign*}
Here, we incorporate our security assumption into the model, marking the $PMK$ as a low-entropy key.

\noindent
\textbf{Property Specification:} We will verify the secrecy of $PTK$ to analyze whether attackers can exploit a low-entropy $PMK$ to compromise the security of WPA2-PSK.
We introduce a fact, \TFact{SuppInsPTK}, indicating that the supplicant has installed a $PTK$.
The definition of the secrecy of the $PTK$ is as follows:
\begin{flalign*}
    & \hspace{0em} \TNot \, (\TEx \;\, \text{\Tterm{suppID  PTK \#i \#j.}} \;\, &\\
    & \hspace{1em} \text{\TFact{K(PTK)@\#i}} \; \land \; \text{\TFact{SuppInsPTK(suppID, PTK)@\#j}})
\end{flalign*}
This formula indicates that the attacker can't know a $PTK$ while the supplicant has installed it.

\subsection{Paring Protocol of BLE}
\begin{figure}[t]
    \centering
    \includegraphics[width=\columnwidth]{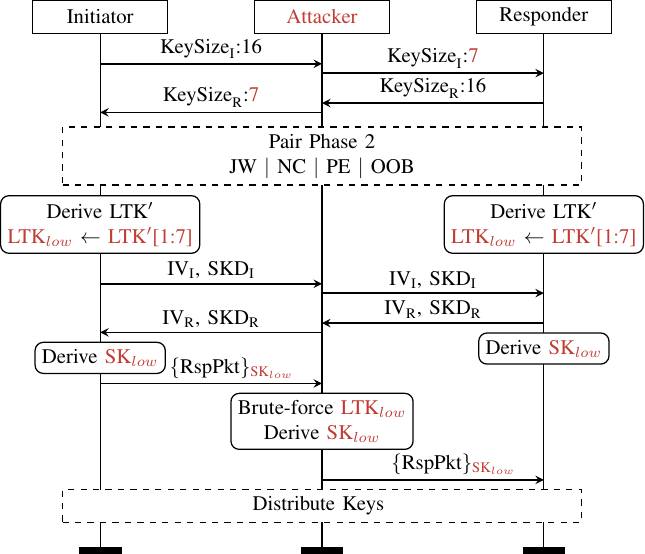} 
    \caption{Attack flow of KNOB on BLE pairing}
    \label{fig:ble-knob}
\end{figure}

In the BLE pairing protocol, there is a KNOB attack\cite{antonioli2019knob}. 
Through this attack, an attacker can downgrade the negotiated key length between the pairing parties, thereby enabling the attacker to compromise the long-term key through brute force.
As illustrated in \autoref{fig:ble-knob}, the attacker manipulates the $LTK$ length field within the pairing features exchanged between the Initiator and Responder, leading to the negotiation of a shorter $LTK$.
This alteration substantially reduces the complexity of brute-forcing $LTK$.

\noindent
\textbf{Security Assumption:} 
The Bluetooth core specification\cite{bluetooth6} states that Bluetooth Low Energy devices can support key lengths of 7 to 16 bytes. For convenience, we assume that the devices involved in pairing support key lengths of 8 to 16 bytes. In this scenario, without the involvement of an attacker, these devices will negotiate a full-length LTK.
Additionally, we assume that the attacker can brute-force an 8-byte key, meaning the attacker can quickly traverse the key space of 8 bytes or fewer.
\BlueQuote{[Vol 3, Part H, 2.3.4]}{ 
    The maximum and minimum encryption key length parameters shall be between 7 octets (56 bits) and 16 octets (128 bits), in 1 octet (8~bit) steps.
}

\noindent
\textbf{Protocol Modeling:}
\cite{shi2023blesc} has developed a comprehensive formal model for the BLE-SC pairing protocol, which includes all related models specified in the standards.
According to the Bluetooth Specification, the final process in $LTK$ derivation involves truncating the $LTK$ to the negotiated key length.
For instance, if the negotiated key length is 7 bytes, only 7 bytes of the $LTK$ will remain as valid data.
Given the computational capabilities of an adversary, shorter $LTK$s are susceptible to being compromised, whereas a full-length $LTK$ remains secure against brute-force attacks.
Existing research defines a rule that provides the attacker with the ability to access low-entropy keys:
\begin{flalign*}
    \text{\ruleNameText{Rule R}}    & \text{\ruleNameText{evealLTK}} :                                   & \\
                            \TLetS  & \text{\Tterm{$LTK$ = resize(\_, "Low Entropy")}}       \\
     \TInS \TFacts{                 & \text{\TFactTerm{In}{senc(m, $LTK$)}}} \TEmpActFact{} \TFacts{ \text{\TFactTerm{Out}{$LTK$}}}             
\end{flalign*}
Here \funcText{resize} truncates the key given in its first argument to the length given in its
second. This rule gives the attacker a low-entropy $LTK$ in principle, but not in the protocol as
deployed: paired devices derive a session key (SK) from the $LTK$ and encrypt with the SK, so a
Dolev-Yao attacker never sees a ciphertext under the $LTK$ and the rule never fires. That model
finds the KNOB attack only because it lets the $LTK$ protect the traffic directly. We therefore
model the derivation of the SK from the $LTK$ and its use in securing communications.

Furthermore, we modeled the \funcText{resize} function with a finer granularity.
From the attacker's perspective, we define two key lengths, $N_1$ and $N_2$, where $N_1$ denotes the key lengths susceptible to attack, and $N_2$ denotes those remaining secure.
To provide a detailed model of the $LTK$ truncation process, we introduce two functions, \funcText{left} and \funcText{right}, which respectively denote the left and right halves of the key.
The \funcText{resize} function satisfies the following equations:
\begin{flalign*}
    \text{\Tterm{resize(k, $N_1$)}} & \text{ = \Tterm{left(k)}}    & \\
    \text{\Tterm{resize(k, $N_2$)}} & \text{ = \TPairN{left(k), right(k)}}    
\end{flalign*}

Taking into account the computational capabilities of an adversary, both \Tterm{left(LTK')} and \Tterm{right(LTK')} are vulnerable from the attacker's perspective
Therefore, in our model, both of these $LTK$ parts are marked as low-entropy keys.
The $LTK$ truncation process is modeled as follows:
\begin{flalign*}
    \text{\ruleNameText{Rule R}}    & \text{\ruleNameText{esizeLTK}} :                                   & \\
        \TLetS                  & \text{\Tterm{LHalf, RHalf := `L`}}                     \\
                                & \text{\Tterm{$LTK$ = resize(LTK', KS)}}             \\
                                & \text{\Tterm{LHalf = left(LTK')}}                  \\
                                & \text{\Tterm{RHalf = right(LTK')}}                 \\
        \TInS \TFacts{          & \text{\TFactTerm{InitState}{"CALCING", \TPairN{LTK', KS}}}}   \\
        \TEmpActFact{ } \hspace{0.5em} \TFacts{ & \text{\TFactTerm{InitState}{"CALCED", $LTK$}}}                  
\end{flalign*}
In this rule, \Tterm{KS} denotes the negotiated key length, while \Tterm{LHalf} and \Tterm{RHalf} represent the left and right halves of \Tterm{LTK'} respectively.
They are labeled as low-entropy keys.

\noindent
\textbf{Property Specification:}
Our focus is on analyzing whether an attacker can leverage computational capabilities to compromise keys, thus we emphasize the study of key confidentiality within the protocol.
We define two facts, \TFact{LtkNegotiated} and \TFact{SkDistributed}, representing the successful negotiation of $LTK$ and the successful distribution of the SK, respectively.
The secrecy of $LTK$ for the Initiator and Responder is defined as follows:
\begin{flalign*}
    & \hspace{0em} \TAll \;\, \text{\Tterm{Addr LTK \#i.}} \;\, &\\
    & \hspace{1em} \text{\TFact{LtkNegotiated(Role, Addr, LTK)@\#i}} \\
    & \hspace{0em} \implies \TNot (\TEx \;\, \text{\TLemma{\#j.}} \;\, \text{\TFact{K(LTK)@\#j}} ) 
\end{flalign*}
where \TFact{Role} could be ``Initiator'' or ``Responder''.
This formula indicates that it is impossible for the attacker to know the $LTK$ once the initiator or responder has successfully negotiated it.
Similarly, the Secrecy of $SK$ means the attacker cannot know the $SK$ while the initiator or responder has successfully negotiated it, shown as follows:
\begin{flalign*}
    & \hspace{0em} \TAll \;\, \text{\Tterm{Addr SK \#i.}} \;\, &\\
    & \hspace{1em} \text{\TFact{SkDistributed(Role, Addr, SK)@\#i}} \\
    & \hspace{0em} \implies \TNot (\TEx \;\, \text{\TLemma{\#j.}} \;\, \text{\TFact{K(SK)@\#j}} ) 
\end{flalign*}

\subsection{Session Establishment of BC} \label{sec:bc-devices}

\begin{figure}[t]
    \centering
    \includegraphics[width=0.962747\columnwidth]{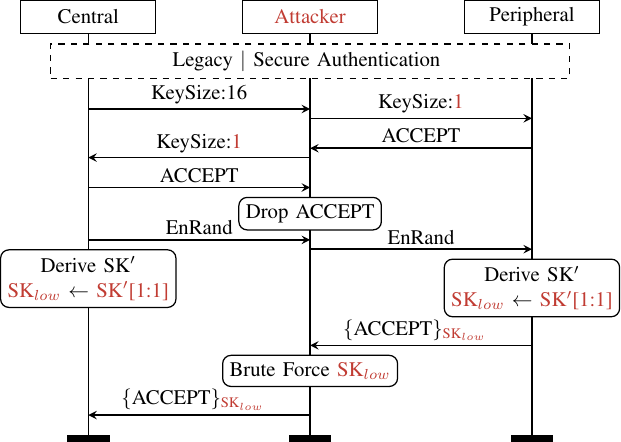} 
    \caption{Attack flow of KNOB on BC session establishment}
    \label{fig:bc-knob}
\end{figure}

BC devices are  also vulnerable to KNOB attacks due to their negotiation of session key lengths during the establishment of sessions.
The attacker compromises the security by altering the negotiated session key length between the central device and the edge device, resulting in a shorter session key. This reduction in key length significantly eases the difficulty for the attacker to brute-force the session key.
\autoref{fig:bc-knob} illustrates the process of a KNOB attack in BC devices.

\noindent
\textbf{Security Assumption:} 
The Bluetooth Core Specification is inconsistent about the minimum key size. Since
version~5.2~\cite{bluetooth52} it states that \Tterm{L\_max} \emph{shall} and \Tterm{L\_min}
\emph{should} be at least 7~octets~[Vol~2, Part~C, 4.2.5.2], where \emph{shall} is mandatory and
\emph{should} is a recommendation, while a second passage requires only $1 \leq
\text{\Tterm{L\_max}} \leq 16$~[Vol~2, Part~H, 4.1]~\cite{bluetooth6}.
Version~5.1~\cite{bluetooth51} and earlier allow a minimum of 1~byte.
We therefore assume three types of key length ranges accepted by BC devices:
\begin{itemize}[leftmargin=*]
    \item \textbf{Secure Keysize:} The BC device only accepts keys of 16 bytes and rejects any keys shorter than 16 bytes.
    \item \textbf{Enhancing Keysize:} The BC device accepts keys ranging from 7 to 16 bytes, but prefers 16-byte keys.
    \item \textbf{Legacy Keysize:} The BC device accepts keys ranging from 1 to 16 bytes, but prefers 16-byte keys.
\end{itemize}
Furthermore, devices are classified according to their support for the SC feature into three categories: SC-Only, SC-Legacy, and Legacy-Only.
These categories correspond to devices that only support Secure Connections, those that support both Legacy and Secure Connections, and those limited to Legacy connections, respectively.
\cite{padgette2017guide} recommends that devices adhere to the following guidelines: only use secure connections and only use the maximum key length.
By integrating the SC feature with key support characteristics, six distinct device types emerge.
Throughout the paper we refer to these six combinations using a compact \textsc{Sc}-\textsc{Key} notation: the first component is one of \textbf{ScOn} (Secure-Connections only), \textbf{ScLg} (Secure plus Legacy), or \textbf{LgOn} (Legacy only); the second is one of \textbf{Secure} (16-byte keys only), \textbf{Enhanc} (7--16 byte keys), or \textbf{Legacy} (1--16 byte keys).
For example, \textbf{ScOn-Secure} denotes a device that supports only Secure Connections and accepts only full-length session keys, while \textbf{LgOn-Legacy} denotes a legacy-only device that accepts any key length down to 1 byte.
These abbreviations also index the rows of Table~\ref{tab:bc-session-result}.

We evaluate BC under both graded adversaries of Section~\ref{sec:threat-model}, the semi-attacker
(SA) and the quart-attacker (QA), since the minimum key length can be as low as 1 byte.

\noindent
\textbf{Protocol Modeling:}
We have developed a comprehensive formal model for BC's SC and Legacy session establishment protocols to facilitate the verification of various device types under our specified security assumptions.

Due to the shorter key length supported by BC devices, we introduce more key lengths, $N_1$, $N_2$, $N_3$, $N_4$, and the resize function satisfies the following equation:
\begin{flalign*}
    \text{\Tterm{resize(k, $N_1$)}} &= \text{\Tterm{l(l(k))}}                   & \\
    \text{\Tterm{resize(k, $N_2$)}} &= \text{\TPairN{l(l(k)),r(l(k))}}          \\
    \text{\Tterm{resize(k, $N_3$)}} &= \text{\TPairN{l(l(k)),r(l(k)),l(r(k))}}  \\
    \text{\Tterm{resize(k, $N_4$)}} &= \langle \text{\Tterm{l(l(k)),r(l(k)),} } \\
                                    &\hspace{6.5em} \text{\Tterm{l(r(k)),r(r(k))}}\rangle
\end{flalign*}
Here, \Tterm{l} and \Tterm{r} to represent the left and the right half of the key, respectively.
This modeling strategy involves partitioning the key into $2^n$ segments, continuing until the shortest segment is susceptible to compromise by an attacker.
\begin{flalign*}
    \text{\ruleNameText{Rule R}}    & \text{\ruleNameText{esizeSKQA}}:                                      & \\
    \TLetS                          & \text{\Tterm{l(l(SK')),r(l(SK')) := `L`}} \\
                                    & \text{\Tterm{l(r(SK')),r(r(SK')) := `L`}} \\
                                    & \text{\Tterm{$SK$ = resize(SK', KS)}}   \\
    \TInS \TFacts{              & \text{\TFactTerm{State}{"SK CALC", \TPairN{SK', KS}}}}             \\
    \TEmpActFact{} \TFacts{     & \text{\TFactTerm{State}{"SK CALCED", $SK$}}}                 
\end{flalign*}
Within the attacker model for SA, we consider keys shorter than half the length to be vulnerable by attackers, with corresponding model rules consistent with those in BLE. 
In contrast, under the QA model, only keys shorter than a quarter of the length are considered vulnerable. 
In our rules, we labeled low-entropy keys like the rule above.

\noindent
\textbf{Property Specification:}
In the BC model, the secrecy of the session key is also a critical consideration.
We define the \TFact{SKDistributed} fact to represent the successful distribution of the session key.
The property is specified as follows:
\begin{flalign*}
    & \hspace{0em} \TAll \;\, \text{\Tterm{Addr SK \#i.}} \;\, &\\
    & \hspace{1em} \text{\TFact{SkDistributed(Role, Addr, SK)@\#i}} \\
    & \hspace{0em} \implies \TNot (\TEx \;\, \text{\TLemma{\#j.}} \;\, \text{\TFact{K(SK)@\#j}} ) 
\end{flalign*}
where \TFact{Role} could be either "Central" or "Peripheral".
This formula indicates that the attacker can't the session key while the central or peripheral device has successfully distributed it.

\section{Results} \label{sec:6results}
This section presents the analysis results of the previous three cases and compares them with
models from the literature.
Our experiment was conducted on a server running Ubuntu 22.04 with an AMD 7950X3D CPU and
192GB RAM.
We used \Tamarin{} as the unmodified verification backend, and EntropyVerif operates purely as
a source-to-source front-end that emits standard \Tamarin{} input.
The source models, the generated models and our own runs are in the repository~\cite{toolrepo},
which also ships the container images that pin the toolchains we used.
We will provide a detailed description of the process for the novel attack discovered from the
BC analysis results.

\subsection{WPA2-PSK}

\cite{cremers2020wpa2}'s formal analysis of the WPA2 protocol revealed its vulnerability to Kracks~(Key Reinstallation Attacks)\cite{vanhoef2017krack}.
While the original study primarily focused on the impact of nonce reuse on protocols, we made slight modifications to their formal model to apply our methodology proposed in this paper.
These modifications enable the attacker to compromise low-entropy keys, allowing us to investigate the security vulnerabilities arising from the misuse of common passwords.
This approach provides an understanding of the potential risks associated with low-entropy key
exploitation.

\begin{table}[ht]
    \centering
    \caption{ Analysis Results of WPA2-PSK } \label{tab:4-way-handshake-result}
    \renewcommand{\arraystretch}{1.2} \setlength{\tabcolsep}{8pt}
    \begin{threeparttable}[b]
        \begin{tabular}{lccc}
        \hline
        Model & Executability & PMKID Attack & Kracks \\
        \hline
        \cite{cremers2020wpa2}'s Model & \textcolor{xhs_green}{\ding{52}} & \textcolor{xhs_green}{\ding{52}} & \textcolor{xhs_red}{\ding{56}} \\
        Our Model & \textcolor{xhs_green}{\ding{52}} & \textcolor{xhs_red}{\ding{56}} & \textcolor{xhs_red}{\ding{56}} \\
        \hline
        \end{tabular}
        \begin{tablenotes}
            \item[] \textcolor{xhs_green}{\ding{52}}: the property is satisfied. \hspace{1em} 
                    \textcolor{xhs_red}{\ding{56}}: the property is falsified.
        \end{tablenotes}
    \end{threeparttable}
\end{table}

As shown in \autoref{tab:4-way-handshake-result}, 
our model can discover not only Kracks same as the original one but also the PMKID attack, which was not apparent in the original analysis results.

\begin{table*}[t]
    \centering
    \caption{Analysis Results of BLE Pairing} \label{tab:ble-pairing-result}
    \renewcommand{\arraystretch}{1.2} \setlength{\tabcolsep}{8pt}
    \begin{threeparttable}[b]
        \begin{tabular}{lccccccc}
            \hline  
            Model & Executability & Auth. of Initiator's LTK & Auth. of Responder's LTK & MitM Protection & Secrecy of LTK \\
            \hline  
            Model in \cite{shi2023blesc} + Session & \textcolor{xhs_green}{\ding{52}} & \textcolor{xhs_red}{A1} & \textcolor{xhs_red}{A1} & \textcolor{xhs_red}{A1} & \textcolor{xhs_green}{\ding{52}}  \\
            Our Model + Session & \textcolor{xhs_green}{\ding{52}} & \textcolor{xhs_red}{A2} & \textcolor{xhs_red}{A2} & \textcolor{xhs_red}{A1} & \textcolor{xhs_red}{A2}  \\
            \hline  
        \end{tabular}
        \begin{tablenotes}
            \item[] \textcolor{xhs_red}{A1}: Method Confusion Attack\cite{von2021method} \hspace{1em}
                    \textcolor{xhs_red}{A2}: Key Negotiation of Bluetooth~(KNOB) Attack\cite{antonioli2019knob} \hspace{1em}
                    \textcolor{xhs_green}{\ding{52}}: the property is satisfied. 
        \end{tablenotes}
    \end{threeparttable}
\end{table*}

\subsection{Pairing of BLE}
There are many formal analysis works\cite{ wu2022formal,shi2023blesc, claverie2023tamarin,jangid2023bluepe} on BLE-SC pairing protocols using \Tamarin{}.
Among these, the symbolic model proposed in \cite{shi2023blesc} stands out as the most comprehensive one.
This work introduced several security properties, including four authentication properties, man-in-the-middle~(MitM) protection, LTK confusion protection, and the secrecy of LTK.
However, the approach taken in this work to model brute-force attacks is tricky and imprecise due to the oversimplification of the key transmission process within the pairing protocol.

We corrected their work by modeling the establishment of encrypted sessions and the transmission of keys and then applied our methodology to the complemented model. 
All the analysis results of the two models are shown in \autoref{tab:ble-pairing-result}.
Our model can discover both the  KNOB attack and Method Confusion Attack\cite{von2021method} which is ignored by the complemented version of \cite{shi2023blesc}'s model.

\subsection{Session Establishment of BC}
We performed a study of the session establishment protocol of BC by analyzing several scenarios under the QA and SA models and the results of our analysis are summarized in \autoref{tab:bc-session-result}.

We have found that the security of BC sessions largely depends on the minimum key length supported by peripheral devices. 
As long as this minimum key length falls within the range that an attacker can compromise, the session key of the peripheral device will inevitably be compromised. 
The situation worsens when the peripheral device is compatible with legacy devices: 
an attacker can repeatedly downgrade the peripheral device's key, ultimately recovering the longest session key, even if the key length far exceeds the attacker’s computational capacity.
This attack against BC devices isn't covered by the existing literature, and we term it the Unilateral Multi-Downgrade(UMD) attack.

\begin{table*}[t]
    \centering
    \caption{Analysis Results of BC Session Establishment Protocol} \label{tab:bc-session-result}
    \renewcommand{\arraystretch}{1.2} \setlength{\tabcolsep}{8pt}
    \begin{threeparttable}[b]
    \begin{tabularx}{\textwidth}{c>{\centering\arraybackslash}X>{\centering\arraybackslash}X>{\centering\arraybackslash}X>{\centering\arraybackslash}X>{\centering\arraybackslash}X>{\centering\arraybackslash}X}
    \hline
    Session Scenario & Adversary Model & SecCenSK & SecPerSK & MitMResistance & Time (s) \\ 
    \hline
    \makebox[2.25em][s]{L\hfill g\hfill O\hfill n}\hspace{0.1em}-\hspace{0.1em}\makebox[2.9em][s]{L\hfill e\hfill g\hfill a\hfill c\hfill y}\hspace{0.5em}$\times$\hspace{0.5em}\makebox[2.25em][s]{L\hfill g\hfill O\hfill n}\hspace{0.1em}-\hspace{0.1em}\makebox[2.9em][s]{L\hfill e\hfill g\hfill a\hfill c\hfill y} & QA & \textcolor{xhs_red}{A1} & \textcolor{xhs_red}{A1} & \textcolor{xhs_red}{A1} & 18.43 \\
    \makebox[2.25em][s]{L\hfill g\hfill O\hfill n}\hspace{0.1em}-\hspace{0.1em}\makebox[2.9em][s]{L\hfill e\hfill g\hfill a\hfill c\hfill y}\hspace{0.5em}$\times$\hspace{0.5em}\makebox[2.25em][s]{L\hfill g\hfill O\hfill n}\hspace{0.1em}-\hspace{0.1em}\makebox[2.9em][s]{E\hfill n\hfill h\hfill a\hfill n\hfill c} & QA & \textcolor{xhs_green}{\ding{52}}  & \textcolor{xhs_green}{\ding{52}}  & \textcolor{xhs_green}{\ding{52}}  & 99172.49 \\
    \makebox[2.25em][s]{L\hfill g\hfill O\hfill n}\hspace{0.1em}-\hspace{0.1em}\makebox[2.9em][s]{E\hfill n\hfill h\hfill a\hfill n\hfill c}\hspace{0.5em}$\times$\hspace{0.5em}\makebox[2.25em][s]{L\hfill g\hfill O\hfill n}\hspace{0.1em}-\hspace{0.1em}\makebox[2.9em][s]{L\hfill e\hfill g\hfill a\hfill c\hfill y} & QA & \textcolor{xhs_red}{A2} & \textcolor{xhs_red}{A2} & \textcolor{xhs_red}{A2} & 32.95 \\
    \makebox[2.25em][s]{L\hfill g\hfill O\hfill n}\hspace{0.1em}-\hspace{0.1em}\makebox[2.9em][s]{S\hfill e\hfill c\hfill u\hfill r\hfill e}\hspace{0.5em}$\times$\hspace{0.5em}\makebox[2.25em][s]{L\hfill g\hfill O\hfill n}\hspace{0.1em}-\hspace{0.1em}\makebox[2.9em][s]{L\hfill e\hfill g\hfill a\hfill c\hfill y} & QA & \textcolor{xhs_red}{A2} & \textcolor{xhs_red}{A2} & \textcolor{xhs_red}{A2} & 81.19 \\
    \makebox[2.25em][s]{S\hfill c\hfill L\hfill g}\hspace{0.1em}-\hspace{0.1em}\makebox[2.9em][s]{S\hfill e\hfill c\hfill u\hfill r\hfill e}\hspace{0.5em}$\times$\hspace{0.5em}\makebox[2.25em][s]{S\hfill c\hfill L\hfill g}\hspace{0.1em}-\hspace{0.1em}\makebox[2.9em][s]{L\hfill e\hfill g\hfill a\hfill c\hfill y} & QA & \textcolor{xhs_red}{A2} & \textcolor{xhs_red}{A2} & \textcolor{xhs_red}{A2} & 150.13 \\
    \makebox[2.25em][s]{S\hfill c\hfill L\hfill g}\hspace{0.1em}-\hspace{0.1em}\makebox[2.9em][s]{S\hfill e\hfill c\hfill u\hfill r\hfill e}\hspace{0.5em}$\times$\hspace{0.5em}\makebox[2.25em][s]{S\hfill c\hfill L\hfill g}\hspace{0.1em}-\hspace{0.1em}\makebox[2.9em][s]{E\hfill n\hfill h\hfill a\hfill n\hfill c} & QA & \textcolor{xhs_green}{\ding{52}}  & \textcolor{xhs_green}{\ding{52}}  & \textcolor{xhs_green}{\ding{52}}  & 83574.33 \\
    \makebox[2.25em][s]{S\hfill c\hfill O\hfill n}\hspace{0.1em}-\hspace{0.1em}\makebox[2.9em][s]{S\hfill e\hfill c\hfill u\hfill r\hfill e}\hspace{0.5em}$\times$\hspace{0.5em}\makebox[2.25em][s]{S\hfill c\hfill O\hfill n}\hspace{0.1em}-\hspace{0.1em}\makebox[2.9em][s]{L\hfill e\hfill g\hfill a\hfill c\hfill y} & QA & \textcolor{xhs_green}{\ding{52}}  & \textcolor{xhs_red}{A1} & \textcolor{xhs_green}{\ding{52}}  & 3993.54 \\
    \makebox[2.25em][s]{S\hfill c\hfill O\hfill n}\hspace{0.1em}-\hspace{0.1em}\makebox[2.9em][s]{S\hfill e\hfill c\hfill u\hfill r\hfill e}\hspace{0.5em}$\times$\hspace{0.5em}\makebox[2.25em][s]{S\hfill c\hfill O\hfill n}\hspace{0.1em}-\hspace{0.1em}\makebox[2.9em][s]{E\hfill n\hfill h\hfill a\hfill n\hfill c} & QA & \textcolor{xhs_green}{\ding{52}}  & \textcolor{xhs_green}{\ding{52}}  & \textcolor{xhs_green}{\ding{52}}  & 13319.46 \\
    \hline
    \makebox[2.25em][s]{L\hfill g\hfill O\hfill n}\hspace{0.1em}-\hspace{0.1em}\makebox[2.9em][s]{S\hfill e\hfill c\hfill u\hfill r\hfill e}\hspace{0.5em}$\times$\hspace{0.5em}\makebox[2.25em][s]{L\hfill g\hfill O\hfill n}\hspace{0.1em}-\hspace{0.1em}\makebox[2.9em][s]{L\hfill e\hfill g\hfill a\hfill c\hfill y} & SA & \textcolor{xhs_red}{A2} & \textcolor{xhs_red}{A2} & \textcolor{xhs_red}{A2} & 82.68 \\
    \makebox[2.25em][s]{S\hfill c\hfill L\hfill g}\hspace{0.1em}-\hspace{0.1em}\makebox[2.9em][s]{S\hfill e\hfill c\hfill u\hfill r\hfill e}\hspace{0.5em}$\times$\hspace{0.5em}\makebox[2.25em][s]{S\hfill c\hfill L\hfill g}\hspace{0.1em}-\hspace{0.1em}\makebox[2.9em][s]{L\hfill e\hfill g\hfill a\hfill c\hfill y} & SA & \textcolor{xhs_red}{A2} & \textcolor{xhs_red}{A2} & \textcolor{xhs_red}{A2} & 67.50 \\
    \makebox[2.25em][s]{S\hfill c\hfill O\hfill n}\hspace{0.1em}-\hspace{0.1em}\makebox[2.9em][s]{S\hfill e\hfill c\hfill u\hfill r\hfill e}\hspace{0.5em}$\times$\hspace{0.5em}\makebox[2.25em][s]{S\hfill c\hfill O\hfill n}\hspace{0.1em}-\hspace{0.1em}\makebox[2.9em][s]{L\hfill e\hfill g\hfill a\hfill c\hfill y} & SA & \textcolor{xhs_green}{\ding{52}}  & \textcolor{xhs_red}{A1} & \textcolor{xhs_green}{\ding{52}}  & 50789.74 \\
    \hline
    \end{tabularx}
    \begin{tablenotes}
        \item[] 
        \begin{itemize}[leftmargin=*]
            \item   \hspace{0.25em} A1: Key Negotiation of Bluetooth~(KNOB) Attack\cite{antonioli2019knob} 
                    \hspace{1em} A2: novel Unilateral Multi-Downgrade Attack 
                    \hspace{1em} \textcolor{xhs_green}{\ding{52}}: the property is satisfied. 
            \item   \hspace{0.25em} ScLg: support both secure and legacy connections.
                    \hspace{1em} ScOn: only support secure connections.
                    \hspace{1em} LgOn: only support legacy connections.
            \item   \hspace{0.25em} Legacy, Enhanc, Secure: key length ranges, see \autoref{sec:bc-devices} for details.
        \end{itemize}
    \end{tablenotes}
    \end{threeparttable}
\end{table*}

\subsection{The Unilateral Multi-Downgrade Attack} \label{sec:umd-attack}
\begin{figure}[t]
    \centering
    \includegraphics[width=0.945285\columnwidth]{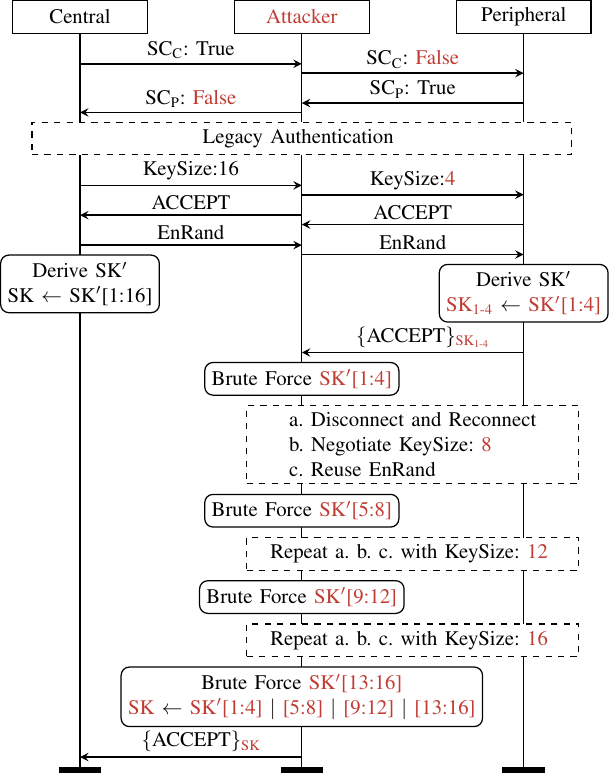} 
    \caption{Attack flow of UMD attack on BC session establishment}
    \label{fig:bc-umd}
\end{figure}

We discover this novel attack where an attacker incrementally downgrades the key length of a device on one side to recover the complete session key.
Given a central device that only accepts 16-byte keys and a peripheral device that supports key lengths from 1 to 16 bytes, their negotiated session key can only be 16 bytes long, making brute-force attacks infeasible.  
Nevertheless, the attacker initiates the attack by downgrading both devices to a Legacy session.
The attacker then induces the peripheral device to use a 4-byte session key, allowing the attacker to brute-force the first 4 bytes of the session key. 
Subsequently, the attacker repeats this process three times, ultimately recovering the complete session key. 
The complete process is depicted in \autoref{fig:bc-umd}.

The UMD attack encompasses several variations, allowing the attacker to freely choose the number of downgrade steps based on their computational resources and the minimum key length supported by the victim device.
If the attacker chooses $n$ downgrade steps, with each step having a consistent interval, the number of computations required by the attacker is $n \times 2^{\frac{128}{n}}$.
When $n$ is 16, the complexity of brute-forcing a 16-byte session key is significantly reduced from $2^{128}$ to $2^{12}$.

The discovery of the UMD attack implies that even if the central device only accepts 16-byte session keys, an attacker can still compromise the session key.
This attack highlights the inherent risks associated with compatibility with legacy devices, as it enables attackers to compromise session keys without eventually downgrading them.

\subsection{Performance Evaluation} \label{sec:performance}

This subsection reports what our translation costs on protocols that contain a low-entropy value
but no reachable brute-force attack, and where the additional proof effort goes.

\noindent\textbf{Protocols with no reachable attack.}
To see what the translation costs when there is nothing to find, we applied EntropyVerif to three
models of the Alipay payment protocols developed by Xiao et al.~\cite{xiao2026payment}: payment
from the mobile application, payment from the mobile web page, and payment by scanning an order
code.
All three carry a user payment password, which is low entropy, but none of them exposes a function
output from which the password could be recomputed.
The change to each model is one label marking the password as low entropy.
Every lemma reaches the same verdict as in the unannotated model and the finished proofs have the
same number of steps, so the analysis outcome does not change. The annotation adds about twelve
seconds to the two simpler models and about four minutes to the order-code model.
Appendix~\ref{app:alipay-time} gives the times.

\noindent\textbf{Where the proof effort goes.}
\Tamarin{} solves constraints in an interdependent way, so its running time cannot be divided
among individual goals. We therefore count the proof steps of each verified BC lemma and assign
each step to the goal it solves. The facts our translation adds take 86.4\% of the steps, and the
goals of the protocol model itself take 54 out of 1{,}779{,}376. Those steps are not new
cryptographic reasoning: a derivation goal asks which terms an observed message was computed from,
and 72\% of them are closed by a rule of the protocol model, so the cost grows with the number of
rules that could have produced a message rather than with the number of low-entropy keys.
Recovering a key is itself almost free, at 0.28\% of the steps. Appendix~\ref{app:breakdown} gives the
method and the per-lemma split.

\noindent\textbf{Human guidance.}
Two kinds of input are supplied by hand, and neither is a lemma.
The first is the low-entropy labeling, which asks for protocol knowledge: for BC, that the session
key is assembled from parts whose lengths the two sides agree on separately.
The second is a heuristic tactic that adjusts goal and branch priorities so that infeasible
branches are closed earlier. It deprioritizes the brute-force goals involving \TFact{MDerive} and
\TFact{EntropyLow}, letting \Tamarin{} analyze them after the others. A tactic restricts no trace
the model admits, so a proof that closes means the same with it as without it.
This strategy is general, although applying it takes minor adjustments for each model: six lines
in each BLE tactic and fourteen in BC. The BC models use tactics alone and carry one, identical in
the SA and QA templates, so all eleven configurations use it, not only the slow ones.

\section{Related Work} \label{sec:7related-work}

We discuss five lines of prior work: empirical studies of low-entropy vulnerabilities, enhanced
attacker models for specific primitives, protocol-specific reveal rules, equivalence-based
analysis of off-line guessing, and probabilistic model checking.

\noindent\textbf{Empirical studies of low-entropy vulnerabilities.}
Early studies~\cite{gong1993protecting,kelsey1997secure} highlighted the profound impact of low-entropy keys on cryptographic-protocol security.
In practice, a long list of severe vulnerabilities has been attributed to low-entropy cryptographic variables, including those disclosed in~\cite{jangid2023bluepe,schneier1998cryptanalysis,ferguson1999cryptographic,gpgcracking} in addition to the KNOB and PMKID attacks examined in this paper.
These studies identify the problem but do not offer an analysis methodology.

\noindent\textbf{Enhanced attacker models for specific primitives.}
Several works extend symbolic tools with enhanced attackers against specific cryptographic primitives, such as weak hash functions~\cite{cheval2023hash}, unsafe Diffie-Hellman groups~\cite{cremers2019prime}, malleable digital signatures~\cite{jackson2019sign}, and AEAD schemes~\cite{cremers2023aead}.
These works target primitive-level weaknesses rather than key-length-induced weaknesses, and typically require modifications to the tool's internal term-rewriting logic.
None of them addresses keys that are weak because of their length, or the decomposition of a
compound key into separately breakable parts.

\noindent\textbf{Protocol-specific reveal rules.}
Shi et al.~\cite{shi2023blesc} proposed an ad-hoc approach to discover brute-force-associated attacks in BLE, where a fixed reveal rule allows the attacker to compromise one particular key.
Claverie et al.~\cite{claverie2023tamarin} and Jangid et al.~\cite{jangid2023bluepe} follow a similar protocol-specific pattern to enable compromise of a single cryptographic variable in BLE pairing.
These approaches are tailored to a known attack and cannot generalize, because (i) the reveal rule hard-codes one specific function symbol, (ii) there is no notion of compound-key decomposition, and (iii) no graded-attacker model constrains which keys are actually within reach of a realistic adversary.
In contrast, our \texttt{MDerive} abstraction is function-symbol agnostic, our three-tier oracle design supports recursive decomposition of compound keys, and our SA/QA attacker model anchors breakability to a concrete bit-level bound.

\noindent\textbf{Equivalence-based analysis of off-line guessing.}
A separate line of work decides whether an attacker can tell a correct guess from a wrong one, by
asking whether two systems that differ only in the guessed value are indistinguishable.
Baudet~\cite{baudet2005guessing} gives the definition in this style, and ProVerif checks it with
diff-equivalence~\cite{blanchet2005equivalences}.
On the \Tamarin{} side, observational equivalence~\cite{basin2015obseq} provides the underlying
method, and Parad\v{z}ik and Derek~\cite{paradzik2022conditional} extend it with conditions to
obtain the first verification of off-line guessing resistance in the multiset rewriting setting,
applied to the EAP-EKE password-authenticated key exchange.
Their question is whether a guess is checkable at all, which is decided from the messages the
protocol emits, and it covers candidate validation that our attacker model leaves out
(Section~\ref{sec:threat-model}).
What the answer does not record is how large the guessed value is. A 128-bit key that is checkable
in principle receives the same verdict as a 4-byte one, so the verdict alone does not separate a
key that a given attacker can enumerate from one that no attacker can.
That separation is what the low-entropy annotation and the graded attacker of
Section~\ref{sec:threat-model} supply.
We also work in \Tamarin{}'s trace mode, so a recovered value is an ordinary term in the
attacker's knowledge and remains usable in later rules and in other sessions, which is what the
UMD attack needs.
The two lines are complementary. An equivalence check tells whether a guess can be confirmed, and
our analysis tells whether confirming it is within a stated attacker's reach.

\noindent\textbf{Probabilistic model checking.}
Siedlecka-Lamch et al.~\cite{siedlecka2016probabilistic} developed a probabilistic model checker for security protocols.
Given a per-key compromise probability, the tool computes a risk tree and outputs a sensitivity assessment.
This line of work is complementary to ours: probabilistic tools quantify leakage risk but cannot produce the kind of concrete attack trace that a symbolic verifier produces, nor can they surface protocol-level multi-step attacks such as UMD.

\noindent\textbf{Summary.}
Across the works above we are not aware of one that combines protocol-agnostic modeling,
compound-key decomposition, and a graded bit-level attacker in a trace-based verifier without
modifying it.
Equivalence-based analyses decide a different and in places broader question, as discussed above,
but do not relate a guess to an attacker's budget.
EntropyVerif occupies that gap, and used it to find a previously unreported protocol-level attack
on Bluetooth Classic.

\section{Discussion and Conclusion} \label{sec:8discussion-conclusion}

\noindent\textbf{Threats to validity.}
Our methodology rests on three explicit assumptions that readers should bear in mind when
interpreting the verification results.
First, our case studies, while substantial, cover three protocol families and six BC device
configurations; generalizing empirical claims about scalability and automation to arbitrary
protocols requires further evaluation.
Second, the attacker of Section~\ref{sec:threat-model} recovers a key only by recomputing a
function forward and matching the result exactly: it does not invert a function over arguments it
does not know, does not accept a candidate on probabilistic evidence, and does not use partial
information about the arguments.
Third, the budget on the nominal key length is chosen by the modeler rather than derived from the
protocol.
If the modeler overstates what the attacker can do and picks a budget looser than that attacker,
the analysis can report attacks that are valid in the symbolic model but infeasible in practice.

\noindent\textbf{Verification effort.}
Although we impose a budget on the value-breaking oracle, verification may still fail to terminate or take an excessively long time on densely interleaved multi-session models; the worst-case BC configuration in our evaluation takes roughly $9.9 \times 10^4$ seconds.
Compared with baseline models that do not carry low-entropy annotations, the analysis cost grows substantially because the relationship-breaking oracle introduces many new derivation branches.
In practice, the BC and BLE models rely on hand-written \Tamarin{} tactics that reorder goals so
that infeasible branches close earlier, and use no auxiliary lemmas. Section~\ref{sec:6results}
describes this guidance.

\noindent\textbf{Implications for the Bluetooth Classic specification.}
The UMD attack is not an implementation bug but a weakness in the protocol design: as long as two endpoints can repeatedly renegotiate the session-key length asymmetrically, and as long as the resulting session keys remain derivable fragments of a shared long-term secret, the attack complexity $C(n) = n \cdot 2^{128/n}$ applies regardless of which device vendor is in use.
Based on the attack, we argue that the Bluetooth SIG should consider the following revisions.
(i) Impose a \emph{hard floor} on the session-key length in the core specification (e.g., 16~bytes for Secure, and at least 7~bytes for Enhanced devices communicating with modern peers), so that the $n = 16$ optimum of $C(n)$ becomes unreachable.
(ii) Require \emph{bilateral confirmation} of any key-length downgrade, so that an attacker coercing one side into a short-key session cannot proceed without simultaneously coercing the other.
(iii) Bind short-key negotiations to the pair of \texttt{BDADDR}s that initiated the long-term key, so that fragments obtained in one session cannot be recombined into a key that targets an unrelated session.

\noindent\textbf{Scope of the methodology.}
Our methodology currently does not model private (secret) functions, because the prerequisites for brute-force attacks are defined in terms of publicly-known cryptographic operators.
When a real protocol uses a private function, one can often replace it with a combination of public functions and private variables, and the resulting model stays within the attacker model of Section~\ref{sec:threat-model}; we leave native support for private functions to future work.
Similarly, our modeling focuses on brute-force attacks whose cost depends solely on key length and does not cover side-channel-assisted or differential attacks whose cost depends on implementation-level details.

\noindent\textbf{Conclusion.}
We proposed the first protocol-agnostic symbolic methodology for analyzing protocols that tolerate
low-entropy keys, built on the \texttt{MDerive} derivation abstraction and a three-tier
breaking-oracle design, and realized as EntropyVerif---a syntax-sugar extension of \Tamarin{} that
emits standard multiset-rewriting rules and leaves the \Tamarin{} backend unmodified.
Our case studies on WPA2-PSK, BLE pairing, and Bluetooth Classic automatically rediscover PMKID, KNOB, and Method Confusion without any protocol-specific reveal rules, and reveal a new protocol-level attack on Bluetooth Classic---the Unilateral Multi-Downgrade (UMD) attack (Section~\ref{sec:umd-attack})---that, via the compound-key decomposition $C(n) = n \cdot 2^{128/n}$ attained at $n=16$, reduces 128-bit session-key recovery by a factor of $2^{116}$.

\section*{Ethics Considerations}

\noindent
\textbf{Responsible disclosure.}
We reported the vulnerability to the Bluetooth Special Interest Group~(SIG) security team on July~15, 2024, 
under the heading of a unilateral key downgrade.
On February~7, 2025 the SIG replied with the results of its internal analysis, considering the
scenario this paper reports to be practical.

UMD is a weakness in how the Bluetooth Classic specification permits key-length negotiation, not a
defect in one product, so we disclosed it to the SIG rather than to individual manufacturers, and
we did not contact any manufacturer directly.
Notification of affected members went through the SIG, which asked for a disclosure timeline on
their behalf and which we authorized to pass our plans on.

\noindent
\textbf{Experimental setup.}
All analysis in this paper is conducted entirely within the symbolic model using \Tamarin{}.
No real Bluetooth devices were involved in the discovery or validation of the UMD attack.
The attack traces are produced by the automated verifier and do not require physical proximity to
any device.
We have therefore not demonstrated the attack against a physical device.

\noindent
\textbf{User-side risk assessment.}
The UMD attack requires a Bluetooth Classic peer that still negotiates the encryption key length
in the legacy manner and accepts a session key of eight octets or fewer. That is the same
precondition as the published key-negotiation downgrade attacks~\cite{antonioli2019knob}, so the
devices at risk are the old ones already at risk from those attacks, not a new population.
After the KNOB disclosure the SIG updated the Core Specification to ``recommend a minimum
encryption key length of 7 octets for BR/EDR connections''~\cite{sig2019knob}. That is a
recommendation rather than a requirement, and it does not exclude the key lengths this attack
needs.
Users can reduce exposure by: (i)~avoiding pairing with untrusted or unknown devices,
(ii)~preferring devices that enforce Secure Connections and maximum key lengths on both sides, and
(iii)~monitoring for unexpected re-pairing events that may indicate an active downgrade attack.

\section*{Acknowledgment}

This research was supported in part by the National Natural Science Foundation of China under Grant Nos.~62625207, 62602479, and 62302271,
and by the China Postdoctoral Science Foundation - Hubei Joint Support Program under Grant No.~2026T077HB.
The  corresponding authors are Jing Chen and Min Shi.

\bibliographystyle{IEEEtran}
\bibliography{library}

\appendices
\section{Verification Time of the Alipay Models}
\label{app:alipay-time}

The three Alipay models of Xiao et al.~\cite{xiao2026payment} contain a low-entropy payment password 
but no function output from which that password could be recomputed, 
so they carry a low-entropy value without carrying a reachable brute-force attack.
They therefore measure what the translation costs when there is nothing for the oracles to find.
Table~\ref{tab:perf-baseline} places each annotated model beside the same model without annotations.
Every lemma reaches the same verdict in both versions and the finished proofs have the same number of steps, 
so the annotation changes the cost of the analysis and not its outcome.

For the application and mobile-web models the annotation adds about twelve seconds, 
under nine percent of the running time.
The order-code model is more expensive in both versions.
Its original tactic leaves attacker-knowledge goals about session values ahead of everything else,
and adding a single deprioritisation line brings the unannotated model to 42\% and the annotated
model to 32\% of its unadjusted time.
We report both variants, and add that line to both versions of the model, so that the pair being
compared differs only in the annotation.
The remaining gap of about four minutes is what the translation costs on this model.

\begin{table}[t]
    \centering
    \caption{Verification time of the Alipay models with and without low-entropy annotations, under \Tamarin{}~1.10.0 and Maude~3.1. ``Adjusted'' adds one deprioritisation line to the tactic of both versions.}
    \label{tab:perf-baseline}
    \small
    \renewcommand{\arraystretch}{1.15}
    \begin{tabularx}{\columnwidth}{@{}>{\raggedright\arraybackslash}X r r r@{}}
    \toprule
    \textbf{Model} & \textbf{Plain (s)} & \textbf{EntropyVerif (s)} & \textbf{$\Delta$ (s)} \\
    \midrule
    Alipay, application          & 134.41 & 146.03 & $+11.62$ \\
    Alipay, mobile web           & 134.10 & 146.16 & $+12.06$ \\
    Alipay, order code           & 456.57 & 1384.68 & $+928.11$ \\
    Alipay, order code, adjusted & 191.08 & 438.45 & $+247.37$ \\
    \bottomrule
    \end{tabularx}
\end{table}

\section{Where the Proof Effort Goes}
\label{app:breakdown}

This appendix gives a detailed breakdown of the proof effort, using the BC case study as the example, 
and reports how much of a completed proof is spent on the goals our translation
introduces.
\Tamarin{} solves constraints in an interdependent way, so its running time cannot be divided
among individual goals, and we decompose the finished proofs instead.
We use BC because it is the only case study with verified all-traces lemmas, 
which are the proofs where the search has to be exhausted and the cost of the translation therefore shows.

\subsection{Method}

\Tamarin{} reports a step count for each lemma.
In the proof script that count corresponds to the \Tterm{case} lines: a \Tterm{solve} on a goal
splits into one or more cases, 
siblings are separated by \Tterm{next}, and the goal is discharged by \Tterm{qed}.
For the \Tterm{SecrecyCentralSK} lemma of one configuration, for instance, the script contains
3292 \Tterm{solve} lines, 3292 \Tterm{qed} lines, 6359 \Tterm{next} lines 
and 9651 \Tterm{case} lines, and \Tamarin{} reports 9653 steps.
Counting cases therefore reproduces the reported figure, and we take one case to be one step.
We assign each step to the goal of the innermost enclosing \Tterm{solve}, and read the kind of that goal from its fact:
\TFact{MDerive0}, \TFact{MDerive1} and \TFact{EntropyLow} are the facts our translation adds, 
\TFact{KU} is attacker knowledge, and the remaining facts belong to the protocol model.
We also record which rule closes each step.

We restrict the count to the lemmas \Tamarin{} verified.
A falsified lemma stops at the first counterexample it finds, so its proof is a single path and
its step count measures how quickly an attack was reached rather than the cost of a search.
A verified all-traces lemma is the opposite case: every branch has to be closed for the property to hold, 
so every case in the script is work that was actually carried out.
The \Tterm{Executability} lemma is also excluded, being an existence property that closes in thirteen to fifteen steps everywhere.
Eleven verified lemmas remain, 1{,}779{,}376 steps in total.

\subsection{Results}

Table~\ref{tab:perf-breakdown-kind} gives the split by kind of goal.
The facts our translation adds account for 86.4\% of the steps, while the goals of the protocol
model itself account for 54 steps out of 1{,}779{,}376.
The tactic sorts the protocol goals ahead of everything else, so they are settled early and the
remainder of the search runs on derivation facts and attacker knowledge.

\begin{table}[t]
    \centering
    \caption{Proof steps of the eleven verified BC lemmas, by kind of goal.}
    \label{tab:perf-breakdown-kind}
    \small
    \renewcommand{\arraystretch}{1.15}
    \begin{tabularx}{\columnwidth}{@{}>{\raggedright\arraybackslash}X r r@{}}
    \toprule
    \textbf{Goal} & \textbf{Steps} & \textbf{Share} \\
    \midrule
    \TFact{MDerive0}          & 1495256 & 84.0\% \\
    \TFact{MDerive1}          &   42504 &  2.4\% \\
    \TFact{KU}                &  241562 & 13.6\% \\
    Protocol state and action &      54 &  0.0\% \\
    \midrule
    Total                     & 1779376 & \\
    \bottomrule
    \end{tabularx}
\end{table}

The \TFact{MDerive0} steps are the largest group, and the rules that close them show what they
are: 72\% are closed by a rule of the BC model, led by \Tterm{PerLegacyEncRsp} with 265{,}936
steps, \Tterm{PerLegacyRecvAuRand} with 153{,}552 and \Tterm{PerSendKeysizeReq} with 139{,}216,
while the remaining 28\% are closed by the two derivation rules the compiler emits.
A \TFact{MDerive0} goal asks which terms an observed message was computed from, and \Tamarin{}
answers it by trying the protocol rules that could have produced that message.
The cost of the translation therefore grows with the number of protocol rules that can produce an
observed message, not with the number of keys marked as low entropy.

The \TFact{KU} steps are spent rebuilding key material.
Constructing the BC key-derivation functions accounts for most of them, \Tterm{c\_h4} alone for
118{,}826 steps, or 49\% of the group, followed by \Tterm{c\_combine} with 69{,}316 steps, or
29\%, which assembles a key from separately recovered parts.
Propagation is minor: the \TFact{MDerive1} facts, which the propagation rule produces, take 42{,}504
steps, one thirty-fifth of the \TFact{MDerive0} total.

The rule that performs the recovery itself, \Tterm{bf\_low\_entropy\_message}, closes 4{,}980
steps, which is 0.28\% of the proofs.

Table~\ref{tab:perf-breakdown} gives the same split per lemma.
Within one configuration the proportions are close to constant across lemmas, so what the
translation costs is a property of the configuration rather than of a particular security
property.

\begin{table}[t]
    \centering
    \caption{Proof steps per verified lemma. MD0, MD1: the derivation facts our translation adds. KU: attacker knowledge. Prot.: goals of the protocol model.}
    \label{tab:perf-breakdown}
    \small
    \renewcommand{\arraystretch}{1.15}
    \begin{tabularx}{\columnwidth}{@{}>{\raggedright\arraybackslash}X l r r r r@{}}
    \toprule
    \textbf{Configuration} & \textbf{Lemma} & \textbf{MD0} & \textbf{MD1} & \textbf{KU} & \textbf{Prot.} \\
    \midrule
    \multirow{3}{*}{\makecell[l]{LgOn-Legacy $\times$\\LgOn-Enhanc, QA}}
        & SecCenSK & 169584 & 5010 &  2254 &  1 \\
        & SecPerSK & 261624 & 7998 &  4682 &  7 \\
        & MitM     & 169584 & 5010 &  2254 &  5 \\
    \midrule
    \multirow{3}{*}{\makecell[l]{ScLg-Secure $\times$\\ScLg-Enhanc, QA}}
        & SecCenSK & 133264 & 3000 & 30028 &  2 \\
        & SecPerSK & 221184 & 4980 & 52477 &  8 \\
        & MitM     & 266528 & 6000 & 60084 & 10 \\
    \midrule
    \multirow{3}{*}{\makecell[l]{ScOn-Secure $\times$\\ScOn-Enhanc, QA}}
        & SecCenSK &  10176 &  324 & 12226 &  1 \\
        & SecPerSK &  15632 &  486 & 19549 &  3 \\
        & MitM     &  10176 &  324 & 12240 &  5 \\
    \midrule
    \multirow{2}{*}{\makecell[l]{ScOn-Secure $\times$\\ScOn-Legacy, QA}}
        & SecCenSK &   7776 &  324 &  1550 &  1 \\
        & MitM     &   7776 &  324 &  1564 &  5 \\
    \midrule
    \multirow{2}{*}{\makecell[l]{ScOn-Secure $\times$\\ScOn-Legacy, SA}}
        & SecCenSK & 110976 & 4362 & 21320 &  1 \\
        & MitM     & 110976 & 4362 & 21334 &  5 \\
    \bottomrule
    \end{tabularx}
\end{table}

\subsection{Summary}

Three things follow.
The goals our translation adds take 86.4\% of the steps of a verified BC proof, and the protocol
model's own goals take 54 of 1{,}779{,}376.
Those steps are not new cryptographic reasoning but a second traversal of the protocol rules,
asking of each observed message which rule could have produced it, which is why 72\% of them are
closed by a protocol rule.
Recovering a key is almost free at 0.28\% of the steps and propagation is cheap at 2.4\%, so what
the analysis pays for is reaching the point at which a recovery becomes possible, and it pays most
where no attack exists and every branch has to be closed.

\end{document}